\documentclass[%
 reprint,
superscriptaddress,
preprintnumbers,
nofootinbib,
nobibnotes,
 amsmath,amssymb,
 aps,
]{revtex4-2}

\usepackage{graphicx}% Include figure files
\usepackage{orcidlink}
\usepackage{dcolumn}% Align table columns on decimal point
\usepackage{bm}% bold math
\usepackage[normalem]{ulem}% strikethrough etc.
\usepackage{hyperref}% add hypertext capabilities
\usepackage{orcidlink}
\graphicspath{{./fig/}{./png/}}
\usepackage{color}
\def\la{\mathrel{\mathchoice {\vcenter{\offinterlineskip\halign{\hfil
$\displaystyle##$\hfil\cr<\cr\sim\cr}}}
{\vcenter{\offinterlineskip\halign{\hfil$\textstyle##$\hfil\cr<\cr\sim\cr}}}
{\vcenter{\offinterlineskip\halign{\hfil$\scriptstyle##$\hfil\cr<\cr\sim\cr}}}
{\vcenter{\offinterlineskip\halign{\hfil$\scriptscriptstyle##$\hfil\cr<\cr\sim\cr}}}}}

\def\RM{\mbox{\rm RM}}
\def\Beq{B_{\rm eq}}
\def\qdiff{q_{\rm diff}}
\def\qball{q_{\rm ballistic}}

\newcommand{\nab}{{\bm{\nabla}}}
\newcommand{\meanA}{{\overline{A}}}
\newcommand{\meanB}{{\overline{B}}}
\newcommand{\meanBB}{{\overline{\bm{B}}}}
\newcommand{\meanJJ}{{\overline{\bm{J}}}}

\newcommand{\Eq}[1]{Eq.~(\ref{#1})}
\newcommand{\Eqs}[2]{Eqs.~(\ref{#1}) and~(\ref{#2})}
\newcommand{\Eqss}[2]{Eqs.~(\ref{#1})--(\ref{#2})}
\newcommand{\Fig}[1]{Fig.~\ref{#1}}

\newcommand{\Tab}[1]{Table~\ref{#1}}

\newcommand{\Sec}[1]{Sec.~\ref{#1}}
\newcommand{\pc}{\,{\rm pc}}
\newcommand{\kpc}{\,{\rm kpc}}
\newcommand{\Mpc}{\,{\rm Mpc}}
\newcommand{\Gyr}{\,{\rm Gyr}}
\newcommand{\kms}{\,{\rm km\,s}^{-1}}
\newcommand{\cm}{\,{\rm cm}}
\newcommand{\m}{\,{\rm m}}
\newcommand{\AU}{\,{\rm AU}}
\newcommand{\s}{\,{\rm s}}
\newcommand{\K}{\,{\rm K}}
\newcommand{\G}{\,{\rm G}}
\newcommand{\eV}{\,{\rm eV}}
\newcommand{\rad}{\,{\rm rad}}
\def\apj{ApJ}

\def\apjs{ApJS}

\def\araa{ARA\&A}

\def\aap{A\&A}
\def\mnras{MNRAS}

\def\nat{NATUR}

\begin{document}

\preprint{NORDITA-2025-055}
\preprint{CERN-TH-2025-219}

\title{%Can astrophysical sources magnetize the voids?
Can galactic magnetic fields diffuse into the voids?}

\author{Oindrila Ghosh}
\affiliation{Oskar Klein Centre for Cosmoparticle Physics, Department of Physics, Stockholm University, AlbaNova, 10691 Stockholm, Sweden}
\email{oindrila.ghosh@fysik.su.se}
%\orcid{0000-0003-2226-0025}
%\orcidlink{0000-0003-2226-0025}
\author{Axel Brandenburg}
\affiliation{Nordita, KTH Royal Institute of Technology and Stockholm University, Hannes Alfv\'ens v\"ag 12, 10691 Stockholm, Sweden}
\affiliation{Oskar Klein Centre for Cosmoparticle Physics, Department of Astronomy, Stockholm University, AlbaNova, 10691 Stockholm, Sweden}
\affiliation{McWilliams Center for Cosmology \& Department of Physics, Carnegie Mellon University, Pittsburgh, PA 15213, USA}
\affiliation{School of Natural Sciences and Medicine, Ilia State University, 3-5 Cholokashvili Avenue, 0194 Tbilisi, Georgia}

\author{Chiara Caprini}
\affiliation{Theoretical Physics Department, CERN, 1211 Geneva 23, Switzerland}
\affiliation{D\'epartement de Physique Th\'eorique and Center for Astroparticle Physics, Universit\'e de Gen\`eve, Quai E. Ansermet 24, 1211 Geneve 4, Switzerland}

\author{Andrii Neronov}
\affiliation{Universit\'e Paris Cit\'e, CNRS, Astroparticule et Cosmologie, 75006 Paris, France}
\affiliation{Laboratory of Astrophysics, \'Ecole Polytechnique F\'ed\'erale de Lausanne, 1015 Lausanne, Switzerland}

\author{Franco Vazza}
\affiliation{Dipartimento di Fisica e Astronomia, Universit\'{a} di Bologna, Via Gobetti 93/2, 40129 Bologna, Italy}

\date{\today}% It is always \today, today,

\begin{abstract}

Cosmic voids are magnetized at the level of at least $10^{-17}\G$ on Mpc scales, as implied by blazar observations.
We show that an electrically conducting
plasma is present in the voids, and that, because of the  
plasma, \emph{diffusion} into the voids of galactic
fields generated by a mean-field dynamo  
is far too slow to explain the present-day void magnetization.
Indeed, we show that even in the presence of turbulence in the voids, dynamo-generated galactic fields diffuse out to a galactocentric radius of only 200--$400\kpc$.
Therefore, it is challenging  to meet the required volume filling-factor of the void magnetic field. 
We conclude that a primordial origin remains the most natural explanation to the space-filling weak fields in voids. 
\end{abstract}

\maketitle

%\tableofcontents
\section{Introduction}
\label{sec:intro}

Magnetic fields in the most tenuous environments of the cosmic web (cosmic voids) are now indirectly constrained at the $\gtrsim 10^{-17}\,\mathrm{G}$ level (for coherence lengths $\gtrsim\mathrm{Mpc}$) by TeV--GeV blazar observations and the nondetection of extended GeV halos \citep{NeronovVovk2010,2025arXiv250622285B,acciari2023lower}. These constraints require that a large \emph{volume fraction} of the line of sight is magnetized: modeling of cascade suppression typically demands filling fractions $\gtrsim 0.6$ for $\gtrsim 10^{-16}\,\mathrm{G}$ fields \citep{Dolag2011,Vazza2017, AlvesBatistaSaveliev2021, 2024ApJ...963..135T}.

Two broad classes of scenarios are considered for the origin of such void magnetic fields. In \emph{primordial} models, magnetic fields are generated in the early Universe (inflationary or phase-transition magnetogenesis) and then processed by magnetohydrodynamic (MHD) decays and cosmic expansion; these scenarios naturally yield spatially extended magnetization that can pervade under-dense regions \citep{DurrerNeronov2013, HoskingSchekochihin2023}. In contrast, \emph{astrophysical} scenarios inject magnetic energy late, via galaxies and active galactic nuclei (AGN) such as winds, jets, outflows, and batteries, resulting in a more patchy cosmic magnetization pattern correlated with matter halos and with a strong dependence on the source population and transport efficiency \citep{KulsrudZweibel2008, Vazza2017}. State-of-the-art cosmological MHD simulations comparing these two families find that primordial seeding tends to produce  larger filling fractions than late, purely astrophysical seeding at fixed observational constraints \citep[e.g.][]{donn09,2015MNRAS.453.3999M,Vazza2017,2021MNRAS.502.5726G,2024ApJ...963..135T}. 

The dynamo mechanism amplifies magnetic fields in galaxies. Such fields are typically quadrupolar in spiral galaxies \cite{parker1971,beck1996,Ntormousi}.
Even if the dynamo in the disk is quadrupolar, that in the galactic halo can be dipolar \cite{SS90,brandenburg1992}.
A recent analysis \citep{GargDurrerSchober2025} 
claims that the ensemble of galactic dipoles alone can produce space-filling pG-level void fields
sufficient to explain the blazar data, even without outflows or turbulence.
This may suggest that dynamo-generated magnetic fields from galaxies in the surrounding clusters can have measurable impact at the center of the voids.
This analysis considers that the late-Universe intergalactic medium (IGM) is a \emph{vacuum} with a \emph{static} superposition of galactic magnetic dipoles. This entails the following: each galaxy with a characteristic $B_0\sim \mu$G magnetic field at a scale $R\sim 10$\,kpc contributes $\sim B_0\,(R/r)^3 \sim 10^{-12}\,$G at $r\sim 1$\,Mpc, before averaging over lines of sight owing to different orientations \citep{Jackson1999}. 

In this article, we argue that the vacuum assumption is not valid on the void length scale because the number of charge-carriers is significant. We show that voids are an electrically conducting medium, which, even preceding the Reionization of the Universe, maintains non-negligible conductivity. Therefore, voids must be treated as plasma. From the plasma physics perspective, \emph{static} magnetic fields from a compact current system could in principle extend into the surrounding medium. 
However, we show that filling previously unmagnetized voids this way within cosmological times is impossible: we demonstrate that even turbulent, as opposed to Ohmic, diffusion, remains extremely inefficient. 

In addition, we show that quadrupole fields have softer fall-off ($\propto r^{-2}$) as opposed to dipolar fields, which spread out from the center of the galaxy as $\propto r^{-3}$. Nevertheless, void magnetization cannot occur regardless of whether the field is dipolar or quadrupolar.

A possible way out might be to admit bulk flow transport (e.g., advection, reconnection in turbulence, outflows) rather than just diffusion. However, the abundance or frequency of energetic astrophysical outflows is highly uncertain, 
limiting the ability of these events to meet observed filling-factors \citep{Aramburo2021,2024ApJ...963..135T}. Therefore, we do not consider them in our analysis. Moreover, the recent discovery of a class of unexpectedly large radio jets  \cite{oei2024black,2025A&A...699A.257A}, might in principle {lead to substantial magnetic pollution,} even though the exact interpretation of these observations remains debated \cite{neronov2025intergalactic}.
We do not consider such extreme objects also on the grounds
that they are sporadic and thus cannot play a pertinent role in magnetizing the voids.

We begin with a discussion of the level of ionization in the voids and the resulting conductivity (\Sec{sec:voidplasma}).
Next, we discuss the radial spreading of magnetic fields generated by a dynamo mechanism both for dipolar and quadrupolar fields (\Sec{sec:dynamo}).
Finally, we discuss the projected rotation measure contribution in both cases (\Sec{sec:rm}), before concluding in \Sec{sec:conc}.

\section{Properties of the void plasma}
\label{sec:voidplasma}

In this section, we examine the nature of the void plasma.
We start with a review of the general properties of plasmas. We then evaluate the conductivity of the Universe over cosmological timescales, taking into account the effect of Reionization. 
This leads us to an estimate of the resistivity in voids today.

\subsection{Plasma conditions from vacuum to voids}

Whether a plasma is collisional depends on the mean free path $\ell_{\text{mfp}}$ of its constituent particles, which in turn depends on the temperature and composition of the plasma, as well as the energy and density of the various constituent particle species, based on which we can establish two different regimes.

When the mean free path $\ell_{\text{mfp}}$ of the particles is significantly smaller than the characteristic system scale $L$, we can consider the plasma to be collisional. In Appendix \ref{app:coll}, we show that this condition is satisfied throughout the thermal history of the Universe on cosmologically relevant scales. In contrast, when the mean free path $\ell_{\text{mfp}}$ is larger than $L$, particles can interact through kinetic effects such as instabilities, Landau damping and turbulence, exemplifying the collisionless regime. For orientation, we summarize the above along with vacuum conditions in \Tab{Regimes}.

\begin{table}[b]\caption{
Conductivities in collisional and collisionless plasmas as well as under vacuum conditions.
}\vspace{12pt}\begin{tabular}{|c|c|c|}
\hline
$\ell_{\text{mfp}} \ll L$  & $\ell_{\text{mfp}} \gg L$ & Vacuum \\
(collisional) & (collisionless) & \\
\cline{1-3}
 Spitzer conductivity & Conductivity from & $\eta \rightarrow \infty$   \\
 $\sigma_{\text{Sp}}$ & Landau damping, instabilities  & $\sigma \rightarrow 0$ \\
\hline
\end{tabular}
\label{Regimes}
\end{table}

The concept of conductivity in a collisional plasma relates to particle scattering. In particular, scattering between electrons, ions, and photons are of importance. Depending on the temperature of the plasma, the scatterings taking place will result in conductivities in the Coulomb or Thomson regimes. A general discussion on Coulomb (Spitzer) and Thomson conductivity can be found below. For collisionless plasmas, on the other hand, wave-particle interactions such as Landau damping and various instabilities such as Buneman or ion-acoustic instabilities lead to an effective conductivity. In addition, turbulence driven in the plasma can contribute to an effective conductivity. Estimates of conductivity  depend on the growth rate of the instabilities directly or through the diffusion coefficient and effective collision frequency.

After reionization, the mean comoving baryon number density is $\bar{n}_b\simeq 2.5\times 10^{-7}\,\mathrm{cm^{-3}}$  \citep{page2003first} at $z=0$; for a fully ionized H/He plasma, this leads to an electron density $\bar{n}_e\approx (0.85\text{--}0.9)\,\bar{n}_b$ \citep{HoskingSchekochihin2023}.
The density contrast in the void is $\Delta\equiv n_e/\bar{n}_e<1$, with typical underdensities implying $n_e\sim 10^{-8}$--$10^{-7}\,\mathrm{cm^{-3}}$ at $z\sim 0$. The exact values of the density depend on redshift and void selection. 

In the following subsection, we discuss general properties of void plasmas and the electric conductivity through cosmic times, ignoring turbulence effects. 

\subsection{Conductivity of the Universe}
\label{subsec:sigmaev}

The Universe can always be treated as a collisional plasma on cosmologically relevant scales \cite{Harrison:1973bt} (see Appendix \ref{app:coll}).
Let us assume that we are in a phase of its evolution in which the only relevant charged particles are nonrelativistic electrons and protons.
Then, the Drude model for the conductivity gives:
\begin{eqnarray}
    \mathbf{j}_{\rm tot}&=&\mathbf{j}_e+\mathbf{j}_p=-en_e \mathbf{v}_e+en_p\mathbf{v}_p= \nonumber \\
   & & e^2 X_e n_b \left(\frac{\tau_e}{m_e}+\frac{\tau_p}{m_p}\right)\mathbf{E},
\end{eqnarray}
where we have set $\mathbf{v}=\pm e (\tau/m) \mathbf{E}$ with $m$ the mass of the particle, $\tau$ its mean-free time between collisions, and $\mathbf{E}$ a test electric field. 
Furthermore, the Universe is neutral, so that $n_e=n_p=X_e n_b$, where $n_b$ is the baryon density, and $X_e=n_e/(n_e+n_{H})$ the time-dependent ionization fraction, where we have neglected Helium \cite{Lewis:2006ym}.
The conductivity is therefore:
\begin{equation}
    \sigma=e^2 X_e n_b \left(\frac{\tau_e}{m_e}+\frac{\tau_p}{m_p}\right).
    \label{eq:sigma}
\end{equation}
For each particle species, one needs to consider the shortest mean-free time between collisions; however, conductivity is governed by the particle species with the longest mean-free time between collisions, weighted by its mass. 

For example, the mean-free time for Coulomb scattering of thermal electrons off protons is (see, e.g., \cite{krall1973principles,Cohen:1950zz})
\begin{equation}
    \tau_{e,C}=\frac{\ell_{e,C}}{\langle v_e\rangle}=\frac{3\,T_b^2}{4\sqrt{2\pi} e^4 \ln \Lambda_\mathrm{c} \,n_b \, X_e(T)}\sqrt{\frac{m_e}{T_b}}\,,
\end{equation}
where $T_b$ denotes the baryon temperature (that we get from \cite{McQuinn:2015icp,Escudero:2023vgv}), while $T$ is the photon temperature (i.e., the one of the Universe),
and $\ln \Lambda_\mathrm{c}$ is the Coulomb logarithm.
Substituting in \Eq{eq:sigma}, accounting also for the protons, this does indeed give the Spitzer conductivity \cite{krall1973principles}  

\begin{equation}
    \sigma_{\rm Sp}=\frac{3\,T_b^{3/2}}{4\sqrt{2\pi} \,e^2 \ln \Lambda_\mathrm{c} \sqrt{m_e}}\,,
    \label{eq:spitzer}
\end{equation}
which depends weakly on $n_b$ through the Coulomb logarithm \cite{richardson20192019}; see Appendix~\ref{app:coulomblog}.
The Spitzer conductivity acquires an additional factor of 2, when considering a thermal electron plasma with a Maxwell distribution.

The mean-free times for electrons due to Thomson and Coulomb scatterings are related via:
\begin{equation}
    \frac{\tau_{e,T}}{\tau_{e,C}}= {\sqrt{2\pi}} 
    \ln \Lambda_\mathrm{c} \,\eta_b\, X_e(T) \left(\frac{m_e}{T_b}\right)^{3/2}\,,
\end{equation}
where we have used $\tau_{e,T}=(\sigma_T n_\gamma)^{-1}$ and $\sigma_T=3m_e^2/(8\pi e^4)$, and 
 $\eta_b=n_b/n_\gamma$ denotes the baryon to photon ratio. 
 For protons, one gets 
$\tau_{p,T}/\tau_{p,C} = (m_p/m_e)^{3/2}(\tau_{e,T}/\tau_{e,C}) $. 
Therefore, Thomson scattering is more efficient for electrons at 
\begin{equation}
    \frac{T_b}{X_e^{2/3}(T)} \geq \left(\frac{\ln \Lambda_\mathrm{c}}{\sqrt{2\pi}} \,\eta_b\right)^{2/3} m_e \simeq  1.9\,{\rm eV},
    \label{eq:elthomson}
\end{equation}
where in the last equality we have substituted $\ln \Lambda_\mathrm{c} =30$ and $\eta_b=6\cdot 10^{-10}$,  
while for protons this becomes $T_b/X_e^{2/3}(T) \geq 1.45 \,{\rm eV}\, (m_p/m_e)\simeq 3291$ eV.
Note that at these high temperatures, $T_b=T$.

At very high temperatures (but still low enough that we are safely in the nonrelativistic limit; otherwise our approach is not valid\footnote{For an evaluation of the conductivity in the very early Universe; see \cite{Ahonen:1996nq,Baym:1997gq,Uchida:2024ude}.}), one expects Thomson scattering to be the most efficient for both particle species, in which case the conductivity is determined by protons that have the longest mean-free time, and is constant (since $X_e\simeq 1$ at high temperature):
\begin{equation}
    \sigma_{p,T} \simeq \frac{3}{8\pi\,e^2}\eta_b\,X_e\,m_p .
\end{equation}
A phase then follows in which the conductivity is still dominated by the protons, although the relevant interaction is Coulomb scattering:
\begin{equation}
    \sigma_{p,C}\simeq\frac{3\,T_b^{3/2}}{4\sqrt{2\pi} \,e^2 \ln \Lambda_\mathrm{c} \sqrt{m_p}}\,.
\end{equation}
This holds until temperatures such that $\tau_{e,T}/m_e\lesssim \tau_{e,C}/m_p$, i.e.,
\begin{equation}
    \frac{T_b}{X_e^{2/3}(T)} \geq \left(\frac{\ln \Lambda_\mathrm{c}}{\sqrt{2\pi}} \,\eta_b \frac{m_e}{m_p}\right)^{2/3} m_p  \simeq  23.3\,{\rm eV}\,.
\end{equation}
From this temperature threshold until the threshold given in \Eq{eq:elthomson}, the conductivity is once again nearly constant, determined by Thomson scattering of electrons, which is the process with the longest mean-free time weighted by the mass:  
\begin{equation}
    \sigma_{e,T} \simeq \frac{3}{8\pi\,e^2}\eta_b\,X_e\,m_e \,.
\end{equation}
However, after the temperature threshold given in \Eq{eq:elthomson}, Coulomb scattering becomes the relevant process for both particle species, and the conductivity is provided by the Spitzer one in \Eq{eq:spitzer}. 

\begin{figure}\begin{center}
\includegraphics[width=0.9\columnwidth]{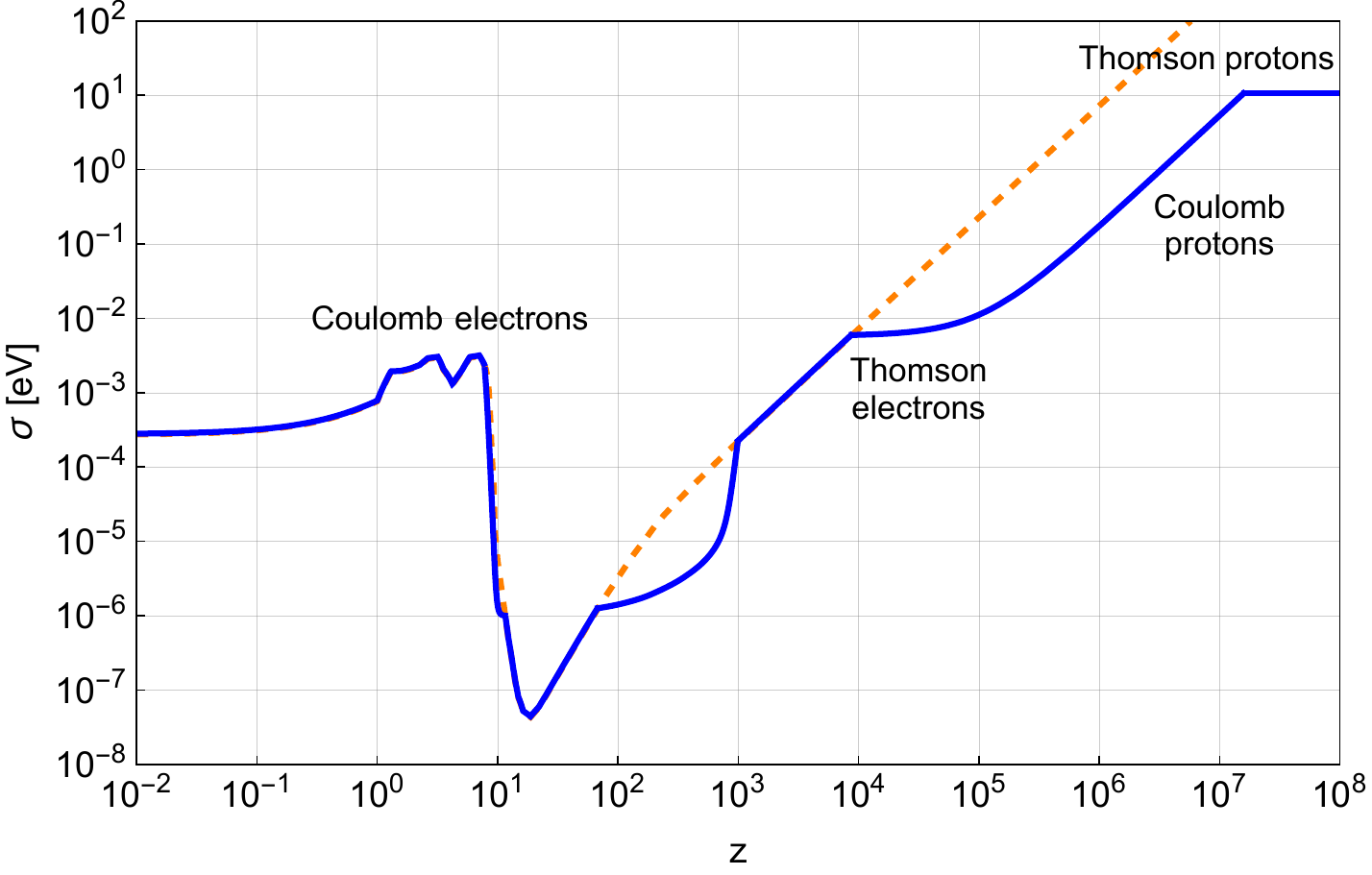}
\includegraphics[width=\columnwidth]{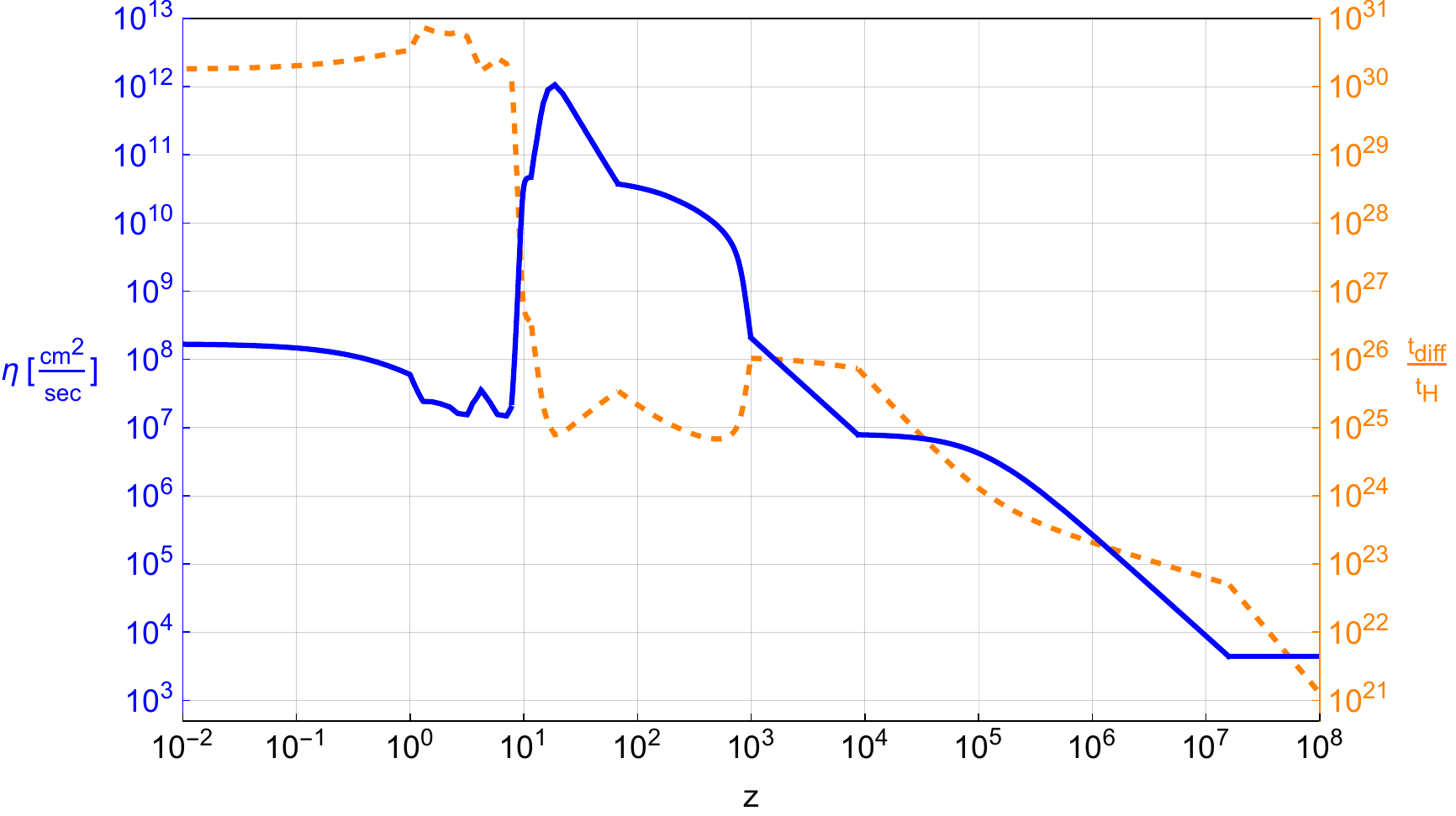}
\end{center}\caption[]{\textit{Upper plot}: conductivity of the Universe (in natural units) as a function of redshift (blue solid line) together with Spitzer conductivity \Eq{eq:spitzer}, i.e. Coulomb scattering of the electrons (orange dashed line). 
\textit{Lower plot}: resistivity $\eta=(\mu_0\sigma)^{-1}$ as a function of redshift (blue solid line), together with the ratio between the diffusion time and the Hubble times, evaluates at the Hubble scale (orange dashed line; see Sec.~\ref{SpreadingByDiffusion}). 
}\label{fig:condu}\end{figure}

The upper panel of Fig.~\ref{fig:condu} shows the Universe conductivity as a function of redshift: the phases derived above are apparent. 
We also show \Eq{eq:spitzer} throughout cosmic time, for comparison. 
The drop in free charges due to Recombination temporarily reduces the conductivity, 
before Reionization increases the ionization fraction $X_e(T)$ again, so that \Eq{eq:spitzer} becomes valid again and the conductivity evolves as $T_b^{3/2}$.  
The effect of the heating of the baryons due to Reionization is also clearly visible, leading to an important increase in the conductivity in the late time Universe. 

The lower panel of Fig.~\ref{fig:condu} shows the magnetic diffusivity.\footnote{Here we use the terms resistivity and magnetic diffusivity interchangeably. We note, however, that in plasma physics, resistivity differs from the magnetic diffusivity by the vacuum permeability. In our case, resistivity and magnetic diffusivity have units of $\cm^2\s^{-1}$.}
In the late Universe, when the conductivity is the Spitzer one of \Eq{eq:spitzer}, the resistivity becomes
\citep{BrandenburgSubramanian2005} 
\begin{equation}
    \eta \simeq 10^7\left(\frac{T_b}{10^4 \mathrm{~K}}\right)^{-3 / 2}\left(\frac{\ln \Lambda_{\mathrm{c}}}{30}\right) \mathrm{cm}^2 \mathrm{~s}^{-1}\,.
    \label{eq:diffspit}
\end{equation}
%\iffalse
%\begin{eqnarray}
%&&    \sigma_{\rm Sp} \simeq 2 \times 10^4\, \frac{T_{\rm eV}^{3/2}}{\ln\Lambda_{\text{c}}}\ ~\mathrm{kg}^{-1} \mathrm{~m}^{-3} \mathrm{~s}^3 \mathrm{~A}^2, \\
%&&\cc{\sigma_{\rm Sp} = 7.8\times 10^{-3} \frac{T_{[\mathrm{K}]}^{3/2}}{\ln\Lambda_{\text{c}}} \frac{\mathrm{~A}^2 \mathrm{~s}^3 }{\mathrm{kg} \mathrm{~m}^{3}}}
%\end{eqnarray}
%\cc{so that for $T\sim 10^4$--$10^5$\,K and $\ln\Lambda\sim 20$ one has $\sigma\simeq 388 - 1.2\times 10^{4} \mathrm{~A}^2 \mathrm{~s}^3 {\mathrm{~kg}^{-1} \mathrm{~m}^{-3}}$,} \textcolor{green}{OG: they are equivalent since 1 eV = $10^4$ K and taking it to power 3/2, we reconcile the difference of six orders of magnitude}
%and the corresponding magnetic diffusivity (in physical units)
%\begin{eqnarray}
%&&\eta = \frac{1}{\mu_0 \sigma_{\rm Sp}} \simeq 42~\frac{\ln\Lambda_{\text{c}}}{T_{[\mathrm{eV}]}^{3/2}} \ \mathrm{m^2\,s^{-1}} ,\\
%&&\cc{\eta = \frac{1}{\mu_0 \sigma_{\rm Sp}}\simeq 10^{8} \frac{T_{[\mathrm{K}]}^{3/2}}{\ln\Lambda_{\text{c}}} \frac{\mathrm{m^2}}{\mathrm{s}}}
%\end{eqnarray}
%\cc{so that for $T\sim 10^4$--$10^5$\,K and $\ln\Lambda\sim 20$ one has $\eta\simeq 0.1 - 5 \mathrm{~m}^{2} \mathrm{~s}^{-1}$,} 
%\fi
For the baryon temperature today, $T_b\simeq 3100\K$,
this provides $\eta\simeq 10^8 \cm^2\s^{-1}$.
Thus, the magnetic Reynolds number $R_m=uL/\eta$ is enormous even for modest intergalactic flows (velocities $u\sim 100\,\mathrm{km\,s^{-1}}$, and length scales $L\sim 1\,\mathrm{Mpc}$),
and field lines are effectively frozen into the plasma on large length scales.
The Universe is therefore a highly conducting, 
%\an{compared to what? You may want first to have the next sentence.. or just skip the end of this sentence.}, 
nearly ideal-MHD medium \citep{Braginskii1965, KulsrudZweibel2008}.

Figure \ref{fig:condu} clearly %demonstrates 
shows
that
%there is a finite conductivity
the conductivity is never zero
% \an{I guess there is no need to demonstrate this.... it is always finite (btw., zero is also "finite" (:-))} 
%\cc{I think we need this sentence because there are people out there ready to believe that voids are vacuum and have zero conductivity}
throughout the Universe, even before Reionization has re-filled the ambient medium with charged particles. 
The cosmological evolution of the conductivity (and the corresponding magnetic diffusivity) implies that it is not accurate to treat the cosmic voids as vacuum on the void scale. 
%\an{it depends on scale. If I take 1 cubic meter of a void, it is vacuum, right? We even do not need to see what is conductivity, because there are just no particles there....} 
We now proceed to estimate how the presence of a finite magnetic diffusivity influences the spreading of dynamo-generated magnetic fields in voids.
%We will demonstrate that they could not reach beyond the outskirts of galaxies even over cosmological timescales.

\section{Growth and evolution of dynamo-generated magnetic field}
\label{sec:dynamo}

\subsection{Galactic dynamo and magnetic field growth}

The galactic magnetic field can be amplified from tiny magnetic seeds via the mean-field dynamo mechanism, converting kinetic energy from the ionized turbulent plasma into large-scale magnetic energy. 
This mechanism can generate the magnetic field in galaxies through turbulence and differential rotation, with the magnetic field undergoing a growth and a saturation phase.
The resulting large-scale configurations can correspond both to quadrupolar (even parity) or dipolar (odd parity) modes \cite{KrauseRadler1980}.
In thin rotating disks, the lowest-order quadrupole mode is favored due to its lower dissipation \cite{ruzmaikin1988} and thus lower critical dynamo threshold.
This implies that the majority of spiral galaxies, which comprise about 60\% of the galactic population, are dominated by even-parity fields \cite{beck1996}.
On the other hand, more spherical systems may favor dipolar configurations \cite{BTK90}. 

Turbulence in galaxies is driven primarily by astrophysical feedback.
Supernova explosions stir the interstellar medium, injecting turbulent kinetic energy on scales well below $100\pc$ \cite{maclow2004,ferriere2001}.
Differential rotation of galactic disks stretches field lines, while cosmic-ray pressure gradients and stellar winds further amplify irregular motions \cite{haverkorn2015,Beck2015}. This combination of shear and helical turbulence underpins the mean-field dynamo.

\subsection{Evolution of the dynamo-generated galactic field}

\subsubsection{Spreading of astrophysical fields through resistivity}
\label{SpreadingByDiffusion}

The question we would like to address first is whether magnetic fields from galaxies can fill voids \emph{without} outflows, turbulence, or large-scale advection,
i.e., via resistive diffusion in a conducting intergalactic plasma. 

The evolution of the magnetic field is governed by the induction equation,
\begin{equation}
\frac{\partial \mathbf{B}}{\partial t} = \nab\times(\mathbf{V}\times\mathbf{B}) - \nab\times(\eta \,\nab\times \mathbf{B}) \,.
\label{eq:dBdt}
\end{equation}
The role of the velocity $\mathbf{V}$ is that it is responsible for the magnetic field generation in the galaxy and the magnetic field redistribution outside the galaxy.
In particular, the velocity can help rearranging the magnetic field into a nearly force-free configuration.
The last term, by contrast, leads to the magnetic diffusion.
The effects of turbulence with velocity $\mathbf{V}$ enter through the first term.
In order to describe turbulence on large scales, it is convenient to consider the averaged equations \citep{KrauseRadler1980,BrandenburgSubramanian2005},
\begin{equation}
\frac{\partial \overline{\mathbf{B}}}{\partial t} = \nab\times(\overline{\mathbf{V}\times\mathbf{B}}) - \nab\times(\eta \,\nab\times \overline{\mathbf{B}}) \,.
\label{eq:dBmdt}
\end{equation}
The difficulty here comes from the nonlinearity, $\overline{\mathbf{V}\times\mathbf{B}}$.
It has contributions both from the mean fields and its fluctuations, $\mathbf{v}=\mathbf{V}-\overline{\mathbf{V}}$ and $\mathbf{b}=\mathbf{B}-\overline{\mathbf{B}}$.
We assume that the definition of averaging obeys the Reynolds rules \citep{KrauseRadler1980}, in which case
\begin{equation}
\overline{\mathbf{V}\times\mathbf{B}}=\overline{\mathbf{V}}\times\overline{\mathbf{B}}+\overline{\mathbf{v}\times\mathbf{b}}.
\end{equation}
%and the remainder,
%\begin{equation}
%\overline{\mathbf{V}}\times\mathbf{B}+\mathbf{V}\times\overline{\mathbf{B}}+%\mathbf{v}\times\mathbf{b}-\overline{\mathbf{v}\times\mathbf{b}},
%\end{equation}
%has only fluctuations and a vanishing average.
In mean-field electrodynamics \citep{KrauseRadler1980}, it is possible to close the equations by expressing $\overline{\mathbf{v}\times\mathbf{b}}$
in terms of the mean fields, e.g.,
\begin{equation}
\overline{\mathbf{v}\times\mathbf{b}}=\alpha\overline{\mathbf{B}}-\eta_\mathrm{turb}\bm{\nabla}\times\overline{\mathbf{B}},
\label{eq:alpha}
\end{equation}
where $\alpha$ is a pseudoscalar related to the kinetic helicity and responsible for large-scale magnetic field generation,
and $\eta_\mathrm{turb}$ is the turbulent magnetic diffusivity.
Ignoring the $\alpha\overline{\mathbf{B}}$ term, since we do not need it for our argument, Eq.~\eqref{eq:dBmdt} becomes
\begin{equation}
\frac{\partial \overline{\mathbf{B}}}{\partial t} = \nab\times(\overline{\mathbf{V}}\times\overline{\mathbf{B}}) - \nab\times[(\eta+\eta_\mathrm{turb}) \,\nab\times \overline{\mathbf{B}}] \,.
\label{eq:dBmdtdet}
\end{equation}
Thus, we see that in the turbulent case, the effective resistivity becomes $\eta_{\rm eff} = \eta+\eta_\mathrm{turb}$.
%Let us begin with the nonturbulent case.

Let us first use \Eq{eq:dBdt} and the results of \Sec{sec:voidplasma} to show that, due to the very high conductivity of the Universe throughout its thermal history, the magnetic diffusivity is utterly negligible on cosmological scales, and therefore ideal MHD is a good approximation to describe the magnetic field dynamics.
Indeed, as we show in the lower plot of Fig.~\ref{fig:condu}, the diffusion 
time over one Hubble distance is much greater than the Hubble time throughout the cosmological evolution.\footnote{Note that this estimate is valid when electrons and protons are no longer relativistic, but the situation is the same at higher energies \cite{Baym:1997gq,Ahonen:1996nq,DurrerNeronov2013}.}
In particular, today it is
\begin{equation}
    \frac{t_{\rm diff}(L_H)}{t_H} = \frac{L_H^2}{\eta t_H} \simeq 3 \times 10^{30}\,.
\end{equation}
 %\an{May be add that this gets to one when $L\sim 10^{-10}L_H$, i.e. in the parsec range in the present-day Universe, it is in this sense that ideal MHD "approximation" is relevant on the voids scale (10 Mpc), right? }
The scale that gets dissipated over one Hubble time today is
\begin{eqnarray}
    L\sim \sqrt{t_H\,\eta}=2\times 10^{-6}\,{\rm pc}
\end{eqnarray}
meaning that the MHD approximation is certainly valid at the void scale. 

We now focus on the case of interest, i.e., dynamo-generated galactic fields. The spreading of the magnetic field is different at early and late times of the dynamo process.
In the kinematic phase, when the magnetic field is still growing exponentially in time proportional to $e^{\gamma t}$,
where $\gamma$ is the growth rate of the dynamo, the magnetic field tends to spread linearly in time \cite{numerical-paper}.
The unmagnetized exterior is separated from the magnetized region around
the galaxy through a front of radius $r_\text{front}$, where the magnetic field falls off exponentially with radius $r$ proportional to $e^{-\kappa r}$ with a suitable coefficient $\kappa$.
This implies that 
\begin{equation}
|\meanBB(r,\theta,t)|\propto e^{\gamma t-\kappa r}=e^{-\kappa(r-c_\mathrm{front}t)},
\label{Bballistic}
\end{equation}
where $c_\mathrm{front}=\gamma/\kappa$ is the front speed.
Because the expansion velocity is constant, we refer to this regime as ballistic.

%We say that it expands ballistically, i.e., with constant expansion velocity.

The front speed depends on the diffusivity regardless of whether it is turbulent or microphysical (Ohmic).
%(Spitzer).
Therefore, we denote the diffusivity in the following as $\eta_{\text{eff}} = \eta + \eta_{\text{turb}}$.
Using $\kappa=(\gamma/\eta_{\text{eff}})^{1/2}$ in \Eq{Bballistic} yields $c_\mathrm{front}=(\qball\gamma\eta_\text{eff})^{1/2}$ for the front speed.
Here, $\qball$ is a coefficient that we shall determine numerically for a specific model.
Thus, the corresponding front radius has the following time dependence \citep{numerical-paper}
\begin{equation}
r_\text{front}^2(t)=\qball\gamma\eta_\text{eff} t^2
\equiv\ell_\text{ballistic}^2.
\label{qball-def}
\end{equation}
Later, when the dynamo has saturated, the magnetic field can still expand diffusively so that
\begin{equation}
r_\text{front}^2(t)=\qdiff\eta_\text{eff} t
\equiv\ell_\text{diff}^2,
\label{qdiff-def}
\end{equation}
where the front radius now only grows like $t^{1/2}$.
Here, $\qdiff$ is again a coefficient that will be determined numerically.

To get an idea about the possible expansion radii, we now adopt some plausible parameters, focusing first on the nonturbulent case, i.e., we set 
$\eta_{\rm eff}=\eta$ (for the turbulent case, see next section).
For the kinematic phase, we must distinguish between the growth rate of the large-scale magnetic field, which can be rather small,
and that of the small-scale magnetic field, which is larger.
%A conservative estimate for the large-scale dynamo is between 2 $\Gyr^{-1}$ \cite{beck1994} and 3 $\Gyr^{-1}$ \cite{kovacs2025halo}.
%AB: there should be no extra space, because \Gyr does already have a space
A conservative estimate for the large-scale dynamo is between $2\Gyr^{-1}$ \cite{beck1994} and $3\Gyr^{-1}$ \cite{kovacs2025halo}.
For comparison, the typical angular velocity of our galaxy is $30\Gyr^{-1}$.
Typical growth rates of the small-scale magnetic field can be of comparable order.
With $\gamma=2\Gyr^{-1}$, after one Hubble time ($t_{\text{H}} \approx 13.8\Gyr \approx 4\times 10^{17}\,\mathrm{s}$), we have $\gamma t_\text{H}=30$, which corresponds to an amplification factor of $e^{30}\approx10^{13}$.
Using $\eta=10^7\cm^2\s^{-1}$ for $T=10^4\K$ in the void, as seen from Eq.~\eqref{eq:diffspit}, this would lead to a ballistically expanding front radius of about $r_{\text{front}} = \ell_{\text{ballistic}} \sim 10^{13}\cm$.
This is $\sqrt{30}\approx5$ times the diffusion length, i.e., $\ell_{\rm ballistic}\approx5\ell_{\rm diff}$, where
\begin{equation}
\ell_{\rm diff}(t) \equiv (\qdiff\eta t)^{1/2} \approx 2\times10^{12}\cm \quad \text{for}\; t \sim t_\mathrm{H},
\end{equation}
with which the front continues to expand after the initial growth phase. Thus, we have $\ell_{\rm diff}\approx0.004\AU$ after one Hubble time if $\qdiff=1$, and five times larger for the ballistic phase when $\qball=1$.
This length is negligible compared to the void size.

Equivalently, the Ohmic diffusion time across a megaparsec is
\begin{equation}
t_{\rm diff} \sim \frac{L^2}{\eta} \gg 10^{30}\ \mathrm{yr}\quad (L=1\,\mathrm{Mpc}) .
\end{equation}
Hence, in a reionized intergalactic plasma, \emph{resistive diffusion alone cannot transport galactic fields into voids on cosmological timescales}. This conclusion is insensitive to the precise void electron density, because $\sigma_{\rm Sp}$ depends only weakly (logarithmically through $\ln\Lambda_\mathrm{c}$) on density \citep{Braginskii1965, KulsrudZweibel2008}. The exact density dependence of $\ln\Lambda_\mathrm{c}$ is discussed in Appendix \ref{app:coulomblog}. The nonturbulent case is demonstrated in the first row of Table \ref{tab:diffusion}. 

\begin{table*}\caption{
Effective diffusivity $\eta_{\text{eff}}$ and diffusion lengths $\ell_{\rm diff}(t_\mathrm{H})$ after one Hubble time
for different combinations of $u_{\text{turb}}$ and $\lambda_{\text{turb}}$ for $\qdiff=1$.
Note that $1\kpc/(\kms)\approx1\Gyr$ and $1\kpc\kms\approx3\times10^{26}\cm^2\s^{-1}$.
}\vspace{12pt}\centerline{\begin{tabular}{lccccccccc}
       & $u_{\text{turb}}$ & $\lambda_{\text{turb}}$ & \multicolumn{2}{c}{ $\eta_{\text{eff}} = \eta + \eta_\text{turb}$ } & $\ell_\text{diff}(t_\mathrm{H})$ & $\ell_\text{ballistic}(t_\mathrm{H})$ & $t_{\text{diff}}/t_{\text{H}}$ \\
       &     $[\kms]$      &        $[\kpc]$         &    $[\kpc\kms]$    &  $[\cm^2\s^{-1}]$  &        $[\Mpc]$                  &            $[\Mpc]$            &    (for $L$ = 10~\text{Mpc}) \\
\hline
\vspace{-3mm}
\\
No turbulence &       --          &            --          &        $3 \times 10^{-20}$       &      $10^{7}$     &         $7 \times 10^{-13}$                     &             $3.5 \times 10^{-12}$  &  $2 \times 10^{26}$ \\
Cosmic web dynamics     &  10 &           1000          &       $3000$       &      $10^{30}$     &         0.2                      &             1   & $         10^{5}$  \\
Outflow      & 300 &            100          &       $10^4$       &  $3\times10^{30}$  &          0.4                     &             2 & 800    \\
\label{Tsummary}\end{tabular}}\label{tab:diffusion}\end{table*}

\subsubsection{Turbulent diffusivity in the void}
\label{sec:etaturb}

The analysis of cosmological simulations in, e.g., \cite{Ryu2008Science,2014MNRAS.445.3706V} indicates that turbulence is generally well developed in clusters and filaments associated with transonic or mildly supersonic flows, while turbulence may be injected in the voids from large-scale structures \citep{Ryu2008Science}. The injection scale associated with cosmic web dynamics can be obtained by considering the curvature radii of cosmological shocks outside clusters, which are of order a few Mpc, implying outer eddy scales $\lambda_{\text{turb}} $ of order $\sim$Mpc. In addition, magnetized outflows from winds, jets and bubbles can introduce turbulence in the voids, however, with typically smaller injection scale of a few $\times 10^2$\,kpc \cite{Aramburo2021, Aramburo2022, bondarenko2022account}.

In a conducting fluid, small‐scale random motions can mix magnetic flux, which leads to an enhanced effective eddy or turbulent diffusivity (see Eq.~\ref{eq:dBmdtdet}).
For nearly isotropic turbulence, $\eta_{\rm eff}$ is well approximated by the mixing length estimate \cite{Mof78}
\begin{equation}
\eta_{\rm turb}\simeq\tfrac{1}{3}\,u_{\text{turb}}\,\lambda_{\text{turb}} \,,
\label{eq:etat}
\end{equation}
where $u_{\text{turb}}$ and $\lambda_{\text{turb}} $ are the rms turbulent speed and outer (energy–containing) scale,
respectively \citep{KrauseRadler1980,BrandenburgSubramanian2005}.
Compressibility, anisotropy, and magnetic feedback can modify the expression somewhat,
but Eq.~(\ref{eq:etat}) remains the standard benchmark \citep{BrandenburgSubramanian2005,Rogachevskii2018}.

If magnetic transport on large scales is modeled as diffusion with coefficient $\eta_{\rm turb}$, Eq. \ref{qdiff-def} at $t = t_{\text{H}}$ can be written as
\begin{equation}
\ell_{\rm diff}(t_\mathrm{H})\sim 2.2\,\qdiff^{1/2}
\left(\frac{u_\text{turb}}{1\kms}\right)^{1/2}
\left(\frac{\ell_\text{turb}}{1\kpc}\right)^{1/2} ~\text{kpc}.
\end{equation}
As we have seen, for $\gamma t_\mathrm{H}=30$, and assuming the prefactors to be unity, the values of $\ell_\text{ballistic}$ are just $\sqrt{30}\approx5$ times larger.
When turbulence is present, $\eta_{\text{eff}} = \eta + \eta_{\text{turb}} \simeq \eta_{\text{turb}}$ (see Table~\ref{tab:diffusion}), and the time to smear a field across a scale $L$ is
\begin{equation}
t_{\rm diff}(L) = \frac{L^2}{\eta_{\text{eff}}}\;\simeq\;\frac{L^2}{\eta_{\rm turb}}\;=\;\frac{3L^2}{u_{\text{turb}} \lambda_{\text{turb}}}\,.
\label{eq:tdiff}
\end{equation}
Eq.~(\ref{eq:etat}) can be combined with physically motivated ranges of values for void/near–void environments. In particular (see also Table~\ref{tab:diffusion}):

\textit{(a) Cosmic web dynamics:} Based on the analysis of the gas flows in state of the art cosmological simulations,
the turbulent velocities in cosmic voids are of order $u_{\text{turb}}\leq 10 \rm ~km/s$ on $\lambda_{\text{turb}}\sim ~\rm Mpc$ scales set by curved shocks \citep{Ryu2008Science},
and the turbulent pressure is $\leq 0.1 $ of the thermal gas pressure there \citep[e.g.][]{iapichino11,wang24,wang24b}. This yields as maximum value 
$\eta_{\rm turb} \sim 10^{30} \cm^2\s^{-1}$,  $\ell_{\rm diff}(t_\mathrm{H})\approx 0.22~{\rm Mpc},\quad t_{\rm diff}(L{=}1~{\rm Mpc})\approx 300~{\rm Gyr}$.
The case of turbulence owing to cosmic web dynamics is listed in the second row of Table~\ref{tab:diffusion}.

\textit{(b) Outflows:} Starburst events in galaxies can impose additional turbulence into the ambient velocity field in the outskirts of galactic halos, as a result of ionized or cold gas biconic outflows, with typical velocities of $\sim 100$--$300 \rm ~km/s$ for nearby galaxies \citep[e.g.][]{2014ApJ...794..156R} and about twice as large in higher redshift galaxies in the $\sim 10^7$--$10^9 M_{\odot}$ range and with redshift $z \sim 4$--$9$ \citep{carniani2024jades}. In some galaxies, the outflow velocity is higher than the escape velocity from the gravitational potential of the host dark matter halo, meaning expelled outflows are able to propagate into the IGM. Numerical and semianalytical methods predict expansion velocities of the same order, and a typical maximum radius of a few $\sim 100 \rm ~kpc$ reached after $\sim 1 \rm ~Gyr$ of expansion \citep[e.g.][]{2018MNRAS.476.1680S,2019MNRAS.490.3234N}.  
Taking a reference value of $u_{\text{turb}}=300 \rm ~ km/s$ and $\lambda_{\text{turb}}=0.1 \rm ~Mpc$, and in the limiting scenario in which the entirety of the outflow velocity is dissipated into turbulence (while in reality part of it gets dissipated via shock heating), we can roughly estimate $\eta_{\rm turb} \sim 3 \times 10^{30}\cm^2\s^{-1}$, $\ell_{\rm diff}(t_H)\approx 0.37\Mpc,\quad t_{\rm diff}(L{=}1\Mpc)\approx 100\Gyr$.
The case of turbulence driven by outflows can be found in the third row of Table \ref{tab:diffusion}.
Even more conservative is the estimate of ionization front-driven turbulence in voids, for which Ref.~\cite{cain2025kiloparsec} quoted timescales of $1\Gyr$ and length scales of $1\kpc$ leading to $\eta_\mathrm{turb}\approx0.3\kpc\kms=10^{26}\cm^2\s^{-1}$. The authors proposed such turbulence as a source of small-scale dynamo action.
%This would be a rather unconventional proposal for void magnetization in our turbulent cases listed in Table~\ref{tab:diffusion}.}
%AB: inserted ", and, if true, would be even more efficient"
This would be a rather unconventional proposal for void magnetization, and, if true, should be even more efficient in the turbulent cases we presented in Table~\ref{tab:diffusion}.

In all cases, it has to be considered that such turbulent motions have to compete with the systematic inflow of gas from voids onto their surrounding halos, which is estimated to have $\sim 150$--$300\kms$ velocity out to $\sim 20$--$30~\rm Mpc$ from halos \citep{2022MNRAS.517.4458M}. In this sense, our analysis is conservative.
%with the tendency of turbulent diffusion to compete against the net global accretion motion from voids to the surrounding cosmic web. 

Even though $\eta_{\rm turb}\gg \eta$ by $\sim\!23$ orders of magnitude, the low turbulent speeds expected in halo outskirts and the large transport distances required mean that \emph{turbulent diffusion alone} still spreads field lines only over $\sim$Mpc scales during a Hubble time, insufficient to magnetize voids of size $\sim$10–50\,Mpc to the high filling fractions implied by $\gamma$–ray cascade constraints. 
%The situation can change in strongly stirred regions (filaments, cluster outskirts), where $u_{\text{turb}}$ is larger and $\lambda_{\text{turb}} $ is . 
%There, turbulent transport can efficiently mix flux locally but filling \emph{void volumes} generically requires advection by large‐scale outflows/jets or a space‐filling primordial seed, not diffusion alone, as can be seen from Table~\ref{tab:diffusion}.
%
In MHD turbulence, magnetic connectivity changes via turbulent reconnection \cite{LV99},
which enables flux to spread and mix at rates controlled by turbulent amplitudes and injection scales rather than by microphysical resistivity \citep{LazarianVishniac1999,ELV2011,Kowal2009}.
In the super-Alfv\'enic or trans-Alfv\'enic limit, this leads to \emph{effective} transport speeds $\sim u_{\text{turb}}$ and hence to $\eta_{\rm turb}$ of Eq.~(\ref{eq:etat});
in sub‐Alfv\'enic regimes transport is reduced by powers of the Alfv\'enic Mach number \citep[][see also reviews cited therein]{LazarianVishniac1999,ELV2011}. The key conclusion above therefore stands: without substantial turbulent stirring in voids, neither Ohmic diffusion nor turbulent diffusion can spread galactic fields across many Mpc within a Hubble time.

%\red{State-of-the-art cosmological simulations have consistently shown that the effect of feedback from galaxies cannot be large enough to magnetize voids entirely,  owing to the low volume-filling fractions of magnetic pollution by all plausible mechanisms of outflows or jets  \cite{2009MNRAS.392.1008D,Vazza2017, Aramburo2021, 2021MNRAS.502.5726G, 2024ApJ...963..135T}.}

%\red{During Reionization, pressure disequilibrium in ionization fronts can drive small-scale turbulent dynamo which can amplify the magnetic field locally around the filaments \cite{cain2025kiloparsec}. Since the volume filling fraction of such turbulence, and by extension of the magnetic fields, remains highly uncertain, we do not consider them in this article.}

\subsection{Simulation results: growth and evolution of dynamo-generated magnetic fields}

\label{subsec:simb}

In vacuum, the far field of a localized source falls as $r^{-3}$
for a dipole and as $r^{-4}$ for a quadrupole, with higher multipoles
decaying even faster \citep{Jackson1999}.
In a conducting medium, however, as we show below, the
quadrupole field falls off only as $r^{-2}$, provided we are still well within
the diffusion radius $\ell_\mathrm{diff}$, up to which the magnetic field has expanded \cite{numerical-paper}.
As we have seen above, this radius is usually well below $1\Mpc$.
Nevertheless, within this radius, the quadrupole field in a conducting plasma is stronger than the dipole field, which still falls off as $r^{-3}$ for $r<\ell_{\rm diff}$. 

\begin{figure}\begin{center}
\includegraphics[width=\columnwidth]{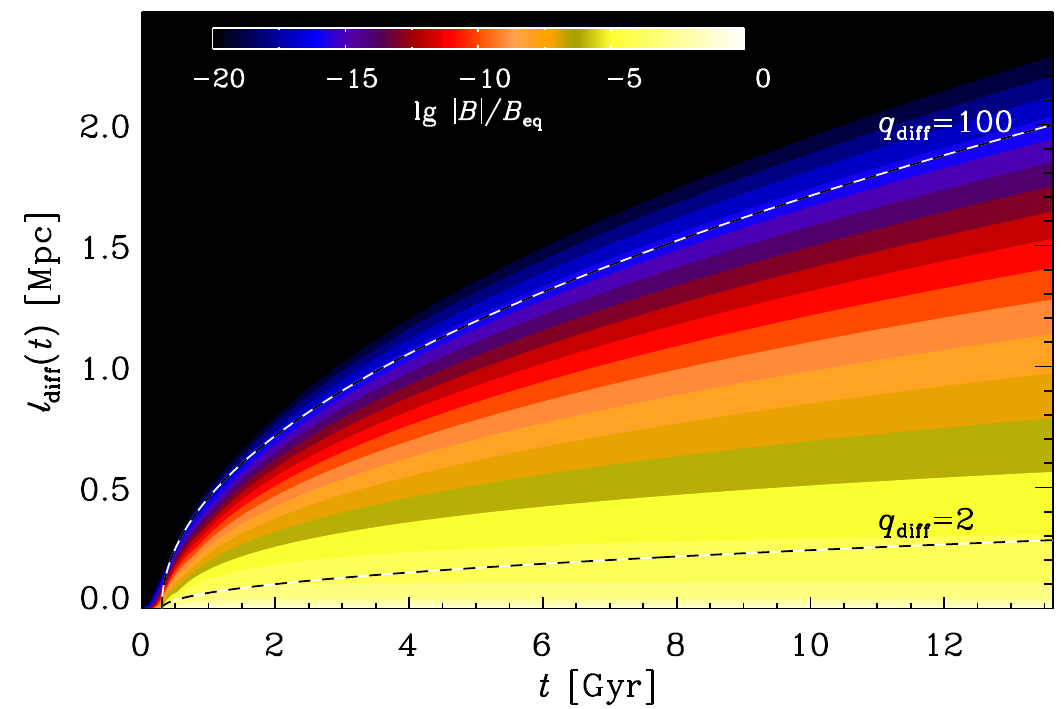}
\includegraphics[width=\columnwidth]{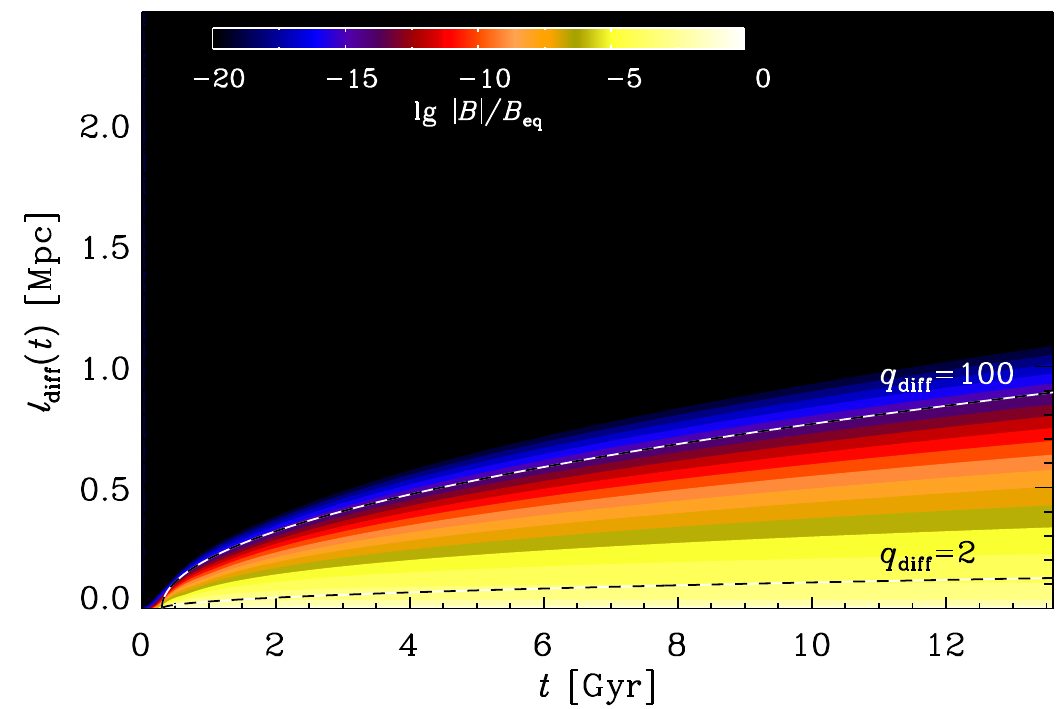}
\end{center}\caption[]{
Logarithm of the quadrupolar $|\meanBB|$ versus time and radius showing the diffusive expansion for
$\eta_{\rm turb}^{\rm ext}=10^{30}\cm^2\s^{-1}$ (top) and $2\times10^{29}\cm^2\s^{-1}$ (bottom).
The lower black dashed and upper white dashed lines denote
$\ell_\text{diff}=[\qdiff\eta_{\rm turb}^{\rm ext}(t-t_\ast)]^{1/2}$
with $\qdiff=2$ and 100, respectively.
}\label{pppb2m_a8192g_rin01_H1f_Q_Pa_rout1e3_alp05_cs5}\end{figure}

To simulate the radial spreading of a dynamo-generated magnetic field
into a possibly poorly conducting exterior, we solve the mean-field
dynamo equations in axisymmetry \citep{KrauseRadler1980}.
In \Eqss{eq:dBmdt}{eq:alpha}, 
the $\alpha$ effect denotes a pseudoscalar related to the kinetic helicity of the turbulence,
and $\eta_{\rm turb}$ is the turbulent magnetic diffusivity defined in \Eq{eq:etat}.
The dynamo is confined to radii $r<R$, where $\alpha\neq0$.
Here, $R$ could be thought of as the typical radius of a galaxy ($R\approx10\kpc$), but we make
no attempt to model any specific aspects of galaxies other than their symmetry about the midplane. 
Notably, in order to optimize the prospects of finding a generic radial scaling behavior, we simulate a spherical domain. Furthermore, we also ignore differential rotation.
The value of $\eta_{\rm turb}$ is finite everywhere, but usually larger in
the exterior 
%($\eta_{\rm turb}=\eta_{\rm turb}^{\rm ext}$), 
than in the interior,
%($\eta_{\rm turb}=\eta_{\rm turb}^{\rm int}$), 
where the dynamo operates.
In the following, we set $\eta_{\rm turb}^{\rm ext}=50\, \eta_{\rm turb}^{\rm int}$ to model a sufficiently diffusive exterior.

The $\alpha$ effect in the interior is 
%assumed to be
%given by $\alpha_0 z/H$, 
proportional to $z/R$,
where $z=r\cos\theta$ is the height above the midplane, and $\theta$ is colatitude.
%For a spherical dynamo domain, we have $H=R$, while for an oblate geometry we have $H<R$.
When the dynamo number $C_\alpha\equiv\alpha_0 R/\eta_{\rm turb}^{\rm int}$
exceeds a certain critical value, 
where the coefficient $\alpha_0$ denotes the strength of the $\alpha$ effect,
there is dynamo action, i.e., the
magnetic field grows exponentially in time starting from a small seed magnetic field.
To be able to reach a steady state, we allow $\alpha$ to be quenched by a factor
$Q(\meanBB)=1/(1+\meanBB^2/\Beq^2)$, where $\Beq$ is the equipartition
magnetic field strength, at which kinetic and magnetic energy densities
are equal.
Therefore, the full dynamo effect is given by $\alpha=\alpha_0 Q(\bar{\bf B})\,z/R$.
We also model the feedback from the large-scale velocity field that is
driven by the Lorentz force of the mean magnetic field $\meanJJ\times\meanBB$.
In the dynamo domain, and for our choice of $\Beq$, this effect is subdominant compared to the effect of $\alpha$ quenching.
In the exterior, however, the Lorentz force helps to make the magnetic field nearly force-free. This is analogous to the magneto-frictional approach used in solar physics \cite{yang1986force}.

For all of our simulations, we use the \textsc{Pencil Code} \cite{PC} using $8192\times32$ mesh points in the radial and latitudinal directions.
We simulate only one quadrant from the pole to the equator, 
for a total volume expanding from $0.1\,R$ to $1000\,R$, setting a dipolar or quadrupolar symmetry condition at the equator. The initial conditions are such that there is a very weak initial magnetic field both in the dynamo region and in the exterior. The initial field in the dynamo determines over how many orders of magnitude the field grows before it saturates.

When $|\meanBB|\la\Beq$, the magnetic field grows exponentially
$\propto e^{\gamma t}$, where $\gamma$ is the growth rate defined in the previous section.
During that stage, as discussed above, the magnetic field spreads radially outward to a radius $\ell_\mathrm{ballistic}$
with a constant velocity $c_\mathrm{front}=(\gamma\,\eta_{\rm turb}^{\rm ext})^{1/2}$. This phase ends at a time $t_\ast$.
When the dynamo saturates, the magnetic field still spreads readily
outward, but at a speed that declines with time as $(t-t_\ast)^{-1/2}$, such that the magnetic
field is confined to a radius $r_\ast(t)=\ell_\mathrm{diff}$, beyond which it declines exponentially.
As shown in \cite{numerical-paper}, the magnetic field can be described well by the formula 
%\cc{isn't there a problem of dimensions here? Maybe $r^{-n}$ needs to be normalized?}
\begin{equation}
\meanB(r,t)=B_0\,\left(\frac{r}{1\kpc}\right)^{-n} \exp\{-\textstyle{\frac{1}{2}}[r/r_\ast(t)]^2\},
\label{eq:Bfit}
\end{equation}
where $\meanB(r,t)$ is the modulus of the latitudinally averaged magnetic field, 
$B_0$ is determined numerically to be of order $\Beq$,
$n$ is an exponent discussed later, and
$r_\ast(t)=[\qdiff\eta_{\rm turb}^{\rm ext} (t-t_\ast)]^{1/2}$ grows with time; see \Eq{qdiff-def}.
This is shown in \Fig{pppb2m_a8192g_rin01_H1f_Q_Pa_rout1e3_alp05_cs5},
where we plot contours of the magnetic field magnitude as a function
of time and radius for two values of $\eta_{\rm turb}^{\rm ext}$
and compare with two choices of $\qdiff$.
We clearly see diffusive expansion with time and that the radial
cut-off decreases with decreasing values of $\eta_{\rm turb}^{\rm ext}$.
This confirms our earlier assertion that in a conducting exterior,
the spreading of magnetic fields is too small to explain the inferred
magnetization in voids.

We now discuss the exponent $n$ in \Eq{eq:Bfit}. 
As demonstrated in \cite{numerical-paper}, our findings are that $n=3$ for a dipolar field, while $n=2$ for a quadrupolar one. The
reason for the unconventional radial $\meanB\propto r^{-2}$ power law for a conducting
exterior is related to the fact that the toroidal field for our quadrupolar mode is of the form $\meanB_\phi(r,\theta)\propto r^{-2}\,P_1^1(\cos\theta)$,
while the toroidal vector potential describing the poloidal field $\meanBB_{\rm pol} = \nabla \times (\meanA_\phi \bm{\hat{\phi}})$ is $\meanA_\phi(r,\theta)\propto r^{-3}\,P_2^1(\cos\theta)$, which is still of the conventional form.
Consequently, the averaged field $\bar B$ 
is dominated by the toroidal component. 
Such a magnetic field is found to emerge naturally in computational domains that are much larger than the dynamo itself.
Here, we recall that we used an outer radius of the domain that is a thousand times larger than the dynamo.
For further detail and additional model results, see Ref.~\citep{numerical-paper}.

Independently, we can use the numerical results to determine the coefficients $\qball$
and $\qdiff$ in \Eqs{qball-def}{qdiff-def}.
For that purpose, it is convenient to determine the instantaneous front
radius $r_\ast(t)$ as a weighted integral,
\begin{equation}
\left.r_\ast(t)=\int r^{n+1} \meanB(r,t)\,d r\right/\!\int r^{n} \meanB(r,t)\,d r.
\label{eq:raverage}
\end{equation}
We can then determine $\qdiff$ from the late-time value of $r_\ast^2/\eta_\mathrm{turb}$.
It turns out that $\qdiff\geq1.7$ for our runs, but it is still slowly
increasing until the end of the run.
It is therefore tempting to assume that the correct value is actually
the same one as for Brownian diffusion, i.e., $\qdiff=2$.
For the ballistic regime, we use a similar procedure as in \Eq{eq:raverage}, except that we choose $n=6$ to account for the steeper radial decay observed in the kinematic regime.
We find a value around $\qball=0.1$ for our spherical models.
This value is rather small, but may also depend on other properties of the model.
For a Cartesian model, by contrast, we typically find a value close to unity \cite{numerical-paper}.

\section{Observability of dynamo-generated fields: Faraday rotation measures}

%\subsection{Observational overview of extragalactic magnetic fields}

\label{sec:rm}

\begin{figure}[b]\begin{center}
\includegraphics[width=\columnwidth]{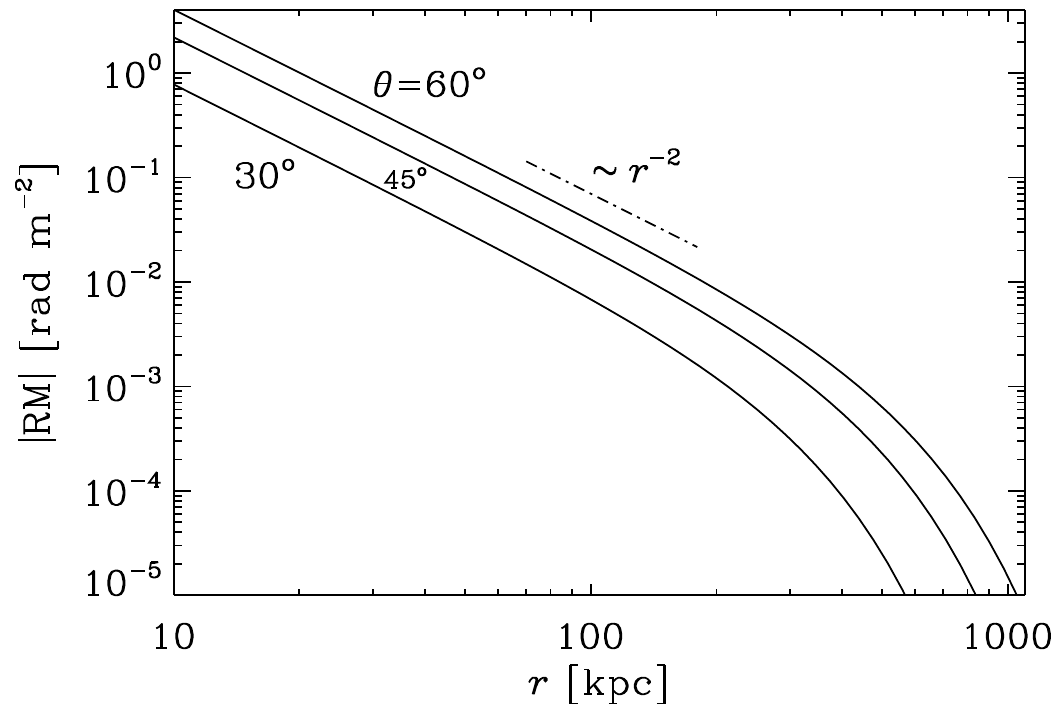}
\end{center}\caption[]{
Radial profiles of RM for the quadrupolar field: we observe a power-law behavior of RM for $r<r_*\simeq 300$ kpc, and a faster decay otherwise.
The dashed-dotted line indicates the $r^{-2}$ scaling.
}\label{prm_Qdiag}\end{figure}

As demonstrated in Sec.~\ref{sec:voidplasma}, dynamo-generated galactic magnetic field cannot spread into the voids within a reasonable timescale (Hubble time). In this section we delve into the potential observable signatures of dipolar and quadrupolar galactic magnetic fields through Faraday rotation measures (RMs), which remain the most direct probe of cosmic magnetic fields.

%Furthermore, we compare our results against the existing simulation studies of astrophysically driven outflows and also against existing observations of extragalactic residual RMs (RRMs). }

%the question remains of how far they reach. 
%n this section we investigate this through simulations. and how they fare against the existing simulation and observational studies on astrophysically driven outflows, and their potential contribution in magnetizing regions of various densities.}

%\red{Observations of the Milky Way’s \emph{Fermi} Bubbles, including linearly polarized radio lobes, confirm large, magnetized outflow structures that advect CRs and power synchrotron, inverse Compton, and hadronic emission across halo scales \citep{Carretti2013Nature,Bordoloi2017ApJ,Heesen2023AandA,Burman2024JCAP,Recchia2021ApJ,Taylor2014PRD}. While cluster‑scale amplification largely erases the memory of whether seed fields were dipolar or quadrupolar in nature \citep{Bertone2006,Donnert2009}, simulations of magnetized bubbles produce fields stronger than the ambient IGMF that can lead to patchy magnetization with low volume-filling fractions \cite{Aramburo2021, Aramburo2023}, with their reach governed by duty cycle and mass loading, giving rise to RMs of a few rad m$^{-2}$ at present day, well below the observed extragalactic residual rotation measures (RRMs).}

On extragalactic scales, Faraday-rotation studies are used to directly probe the extent of magnetization of the circumgalactic medium (CGM). Residual rotation measures (RRMs) show a statistically significant excess for sightlines near the projected minor axes of inclined discs at impact parameters $\lesssim 100$\,kpc, with RRMs of $7.8 \pm 0.9~\mathrm{rad}~\mathrm{m}^{-2}$, and at impact parameters $\gtrsim 100$\,kpc, $4.1 \pm 0.2~\mathrm{rad}~\mathrm{m}^{-2}$ for RRM, leading to an average magnetic field strength of 0.5 $\mu$G along an average sightline length of 100 kpc \citep{Heesen2023AandA}.

\begin{figure}\begin{center}
\includegraphics[width=\columnwidth]{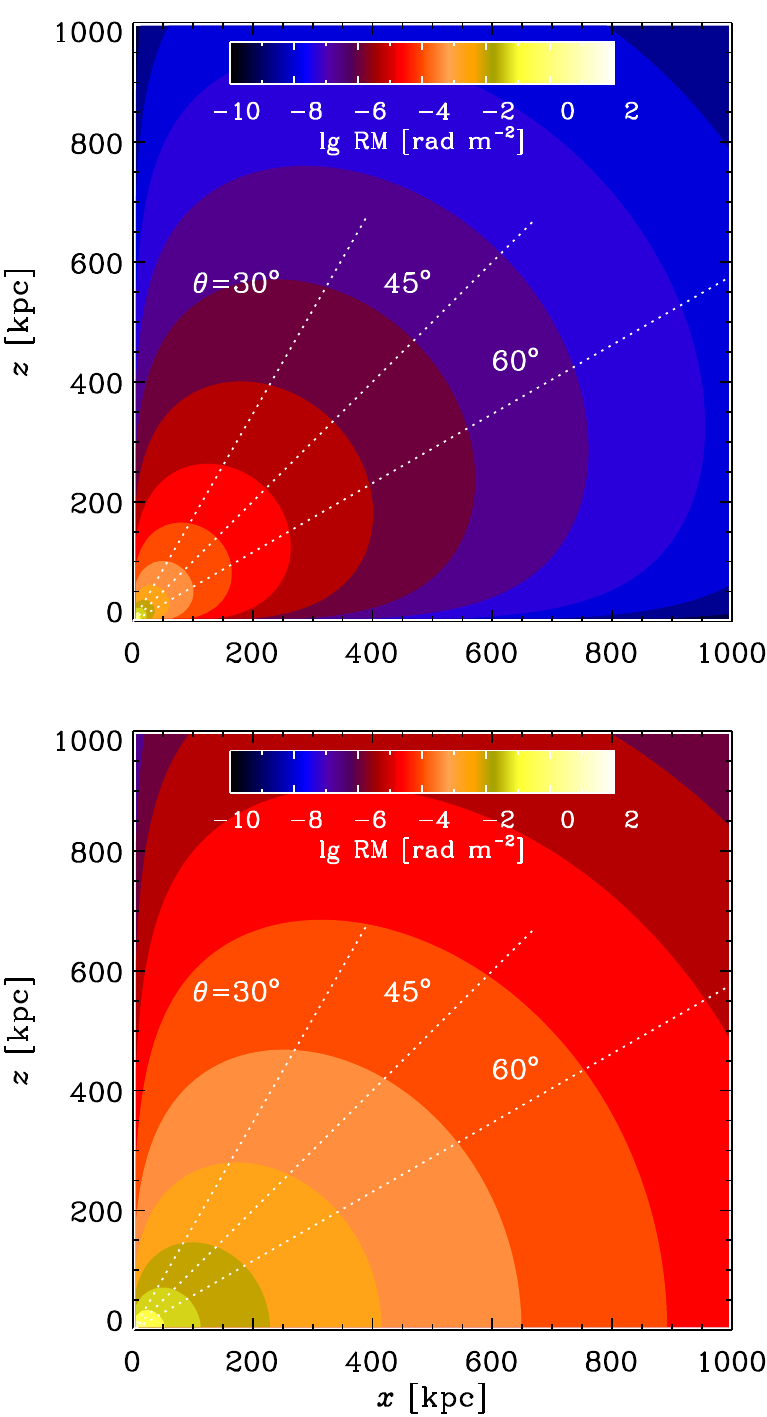}
\end{center}\caption{Plot of RM for a dipolar (upper panel) and quadrupolar (lower panel) field.
The dotted lines denote radial cuts through $\theta=30^\circ$, $45^\circ$, and $60^\circ$: the RM along these lines is shown in  \Fig{prm_Qdiag} for the quadrupolar field.}\label{prm_cut50}\end{figure}

For all galaxies, magnetic fields arising from galactic dynamos will eventually diffuse to a certain galactocentric radius leading to a residual magnetization within a ``magnetosphere'' associated with each galaxy.
In order to address the observability of the diffusion of dynamo-generated galactic magnetic fields, we compute the RM for dipolar and quadrupolar fields. Further details are provided in our companion article \cite{numerical-paper}. 

The power-law dependence of the thermal electron density can be estimated as \cite{ghirardini2019universal}
\begin{equation}
    n_\mathrm{th}=10^{-3}\left(\frac{r}{100~\kpc}\right)^{-s}\cm^{-3}.
    \label{eq:nth}
\end{equation}
Using \Eq{eq:nth}, we can then compute 
\begin{equation}
    \RM=812 \int \left(\frac{n_{\rm th}}{1\cm^{-3}}\right) \left(\frac{B_y}{1~\mu \rm G}\right)\,\left(\frac{dy}{1\kpc}\right) \rad \m^{-2},
\end{equation}
where $B_y$ is the line-of-sight magnetic field given by 
\begin{equation}
B_y=\sin\theta\sin\phi B_r+\cos\theta\sin\phi B_\theta+\cos\phi B_\phi,
\label{eq:By}
\end{equation}
with $\phi$ is the azimuthal angle and $\theta$ is the colatitude.

For power laws of the form $n_\mathrm{th}\propto r^{-s}$, we obtain
$\RM\propto r^{-2-s}$ and $\propto r^{-1-s}$ for the dipolar and quadrupolar solutions, respectively.
In \Fig{prm_Qdiag}, we show radial profiles of RM for the quadrupolar field using \Eq{eq:By}
with $s=1$, which is an approximation in line with the most recent observations obtained by the X-ray eRosita satellite on the Milky Way, providing $s=0.5$--$1.5$ \citep[e.g.][]{2024A&A...681A..78L}. 
Fig.~\ref{prm_Qdiag} clearly shows a radial power-law dependence of the RM. 
For this particular evaluation of the RM, we have fixed $r_*=300$ kpc, the diffusion length expected in one Hubble time for the parameters chosen in the upper panel of \Fig{pppb2m_a8192g_rin01_H1f_Q_Pa_rout1e3_alp05_cs5}. 
The magnetosphere of residual magnetization  extends to a radius of $\sim$ 100 kpc from the galactic center.
An exponential fall-off follows, for radii $r > r_*$.

In \Fig{prm_cut50}, we compare edge-on visualizations for the dipolar and quadrupolar cases.
For the quadrupolar case, RM is the largest at the equator.
This is because in our edge-on view, RM is dominated by the toroidal magnetic field, which is here symmetric about the midplane.
For the dipolar field, the toroidal magnetic field is antisymmetric about the midplane and therefore the RM vanishes there.

%\red{Even though the absolute RM values obtained in our simulation depend strongly on the choice of dynamo model, the values are closer to the observed RM in the circumgalactic medium of nearby galaxies. However, the profile implied by this model are significantly steeper than those observed through LOFAR 
%\cite{Heesen2023AandA} or MeerKAT \cite{2023A&A...678A..56B}.}

The radial RM profiles in Fig.~\ref{prm_Qdiag} and the RM maps in Fig.~\ref{prm_cut50} show values
that drop by two orders of magnitude between $r=10\kpc$ and $100\kpc$.  
For dipolar magnetic fields, they would decay by three orders of magnitude. %Thus, even the unconventionally slow RM decay from our nearly force-free quadrupolar field is significantly faster than the values detected. 
The absolute values of RM in Figs.~\ref{prm_Qdiag} and \ref{prm_cut50} depend heavily on the strength of the galactic dynamo in the model of choice. They are slightly below the current threshold of detection by radio telescopes at low redshifts. 
Moreover, the slope of the reconstructed RM profiles from our simulations is steeper than what currently constrained by LOFAR 
\cite{Heesen2023AandA} or MeerKAT \cite{2023A&A...678A..56B} observations, which may be contaminated by several astrophysical backgrounds, including but not limited to contributions from outflows. However, by comparing with observed rms values of extragalactic RMs from LOFAR Two-Meter Sky Survey (LOTSS), Ref.~\cite{Blunier2024} demonstrated that large-scale structure simulations of outflows such as IllustrisTNG over-predicts the amplitudes \cite{Aramburo2022,aramburo2022revision}.
%Although this may involve contributions from outflows in observed galaxies, among other sources of background, 
 
%Indeed, galaxies, depending on their degree of activity, may be accompanied by some form of outflows such as magnetized bubbles, AGN jets, or starburst winds. Such outflows, as captured in large-scale structure simulations, such as IllustrisTNG \cite{Aramburo2021, Aramburo2023}, can contaminate the rms values of extragalactic RRMs measured by the LOFAR Two-Meter Sky Survey (LOTSS). However, by comparing with observations, Ref.~\cite{Blunier2024} concludes that the amplitudes associated with these events might be overestimated in IllustrisTNG. \red{Moreover, such outflows would produce RMs of a few rad m$^{-2}$ at present day \cite{Aramburo2022,aramburo2022revision}, which are comparable to the RMs presented in Fig. \ref{prm_Qdiag} but well below the observed extragalactic RRMs presented in \cite{Heesen2023AandA,2023A&A...678A..56B}.}

%The size of the magnetosphere is negligibly small compared to the intergalactic distance scales associated with cosmic voids.

%\cite{Blunier2024} %\cite{carretti2025nature}. 

%\fv{I commented out the text after this: it either refers to RM in the intra-group medium or in filaments and has little to do with what we are discussing here}
%The radial dependence of RM owing to individual galaxies has not been observationally explored \og{FV will confirm if such papers exist}, however, median absolute deviation or residual Faraday RM have been studied against splashback radii of galaxy groups \cite{anderson2024probing, carretti2025radio}.

\section{Conclusions}

We have investigated the role of astrophysically generated magnetic fields from galactic mean-field dynamos in magnetizing the voids.
Magnetic fields spread diffusively once the galactic dynamo has saturated to its final field strength.
Prior to that, the magnetic field spreads ballistically, i.e., linearly in time.

By tracing the evolution of the conductivity throughout cosmological time, we evaluate the residual magnetic diffusivity in voids.
Furthermore, we study the additional diffusivity due to the presence of turbulence in voids, which has not been considered in the related recent work \cite{Seller+Sigl2025}. 
Because of the effect of diffusivity, in contrast to the results of \cite{GargDurrerSchober2025} %\red{\sout{and in agreement with \cite{Seller+Sigl2025}}}, 
our main conclusion is that 
galactic magnetic fields cannot penetrate into a distance scale comparable to the void size on a timescale that is less than one Hubble time.
In addition to cosmic web dynamics, which refers to the standard scenario of gas turbulence in the large-scale structure, we explore the role of turbulent diffusion arising from magnetized outflows.
In both cases, the diffusion timescale over the size of the voids far exceeds the Hubble time.

In addition, we performed mean-field MHD simulations of dynamo-generated fields and show that quadrupole magnetic fields falls off as $r^{-2}$ in the conducting void plasma.
This is slower than that in vacuum, where the dipole fields fall off as $r^{-3}$ and quadrupole fields as $r^{-4}$.
Thus, for distances within the diffusion length scale $\ell_{\text{diff}}$, quadrupole fields can survive significantly better than dipole fields.
From the simulations, we clearly see that even the quadrupole field falls off exponentially above length scales of $\mathcal{O}\sim 100\kpc$, which is in line with the estimates of diffusion lengths we show in Table II. This way, we could define a magnetosphere within which the astrophysically generated magnetic field remains largely confined instead of permeating into the void.

Finally, we delved into the observability of such dynamo-generated magnetic fields via Faraday RMs. The details of the dynamo model adopted in the MHD simulation determine the absolute RM values {arising from the magnetic field contained within the magnetosphere.} 
%We find a faster decay of the RM from the galactic magnetosphere with galactocentric radius compared to current LOFAR and MEERKAT observations of extragalactic RRMs. 
%AB: from -> within, compared -> comparable
%We find a faster decay of the RM \red{within} the galactic magnetosphere with a galactocentric radius \red{comparable} to current LOFAR and MEERKAT observations of extragalactic RRMs. 
We find a faster decay of the RM with galactocentric radius, with respect to current LOFAR and MEERKAT observations of extragalactic RRMs.

Upcoming probes, such as rotation measures of fast radio bursts through voids \cite{hackstein2019fast}, combinations of stacked synchrotron maps \cite{chen2015search}, and thermal Sunyaev-Zel'dovich map analyses \cite{minoda2017thermal}, as well as improved pair halo searches associated with blazar cascades can further constrain the filling fractions and coherence lengths of the void magnetic field, shedding light on distinction between the astrophysical and primordial origin scenarios.

\begin{acknowledgments}
CC is grateful to Miguel Escudero for very insightful discussions. 
AB and OG acknowledge valuable discussion with Istvan Pusztai. OG is supported by the Swedish Research Council (Vetenskapsr{\aa}det) under contracts 2022-04283 and 2019-02337 and additionally by the G\"oran Gustafsson Foundation for Research in Natural Sciences and Medicine. This research was supported in part by the Swedish Research Council
(Vetenskapsr{\aa}det) under grant No.\ 2019-04234, the National Science Foundation
under grants No.\ NSF AST-2307698, AST-2408411, and NASA Award 80NSSC22K0825.
We acknowledge the allocation of computing resources provided by the
Swedish National Allocations Committee at the Center for
Parallel Computers at the Royal Institute of Technology in Stockholm. FV has been partially supported by Fondazione Cariplo and Fondazione CDP, through grant n$^\circ$ Rif: 2022-2088 CUP J33C22004310003 for the ``BREAKTHRU'' project. AN is partially supported by the French National Research Agency (ANR) grant ANR-24-CE31-4686.
\end{acknowledgments}

\label{sec:conc}

\bibliography{astrob}
%\nocite{*}
\appendix

\section{Collisionality of the Universe plasma}
\label{app:coll}

\begin{figure}
\includegraphics[width=\columnwidth]{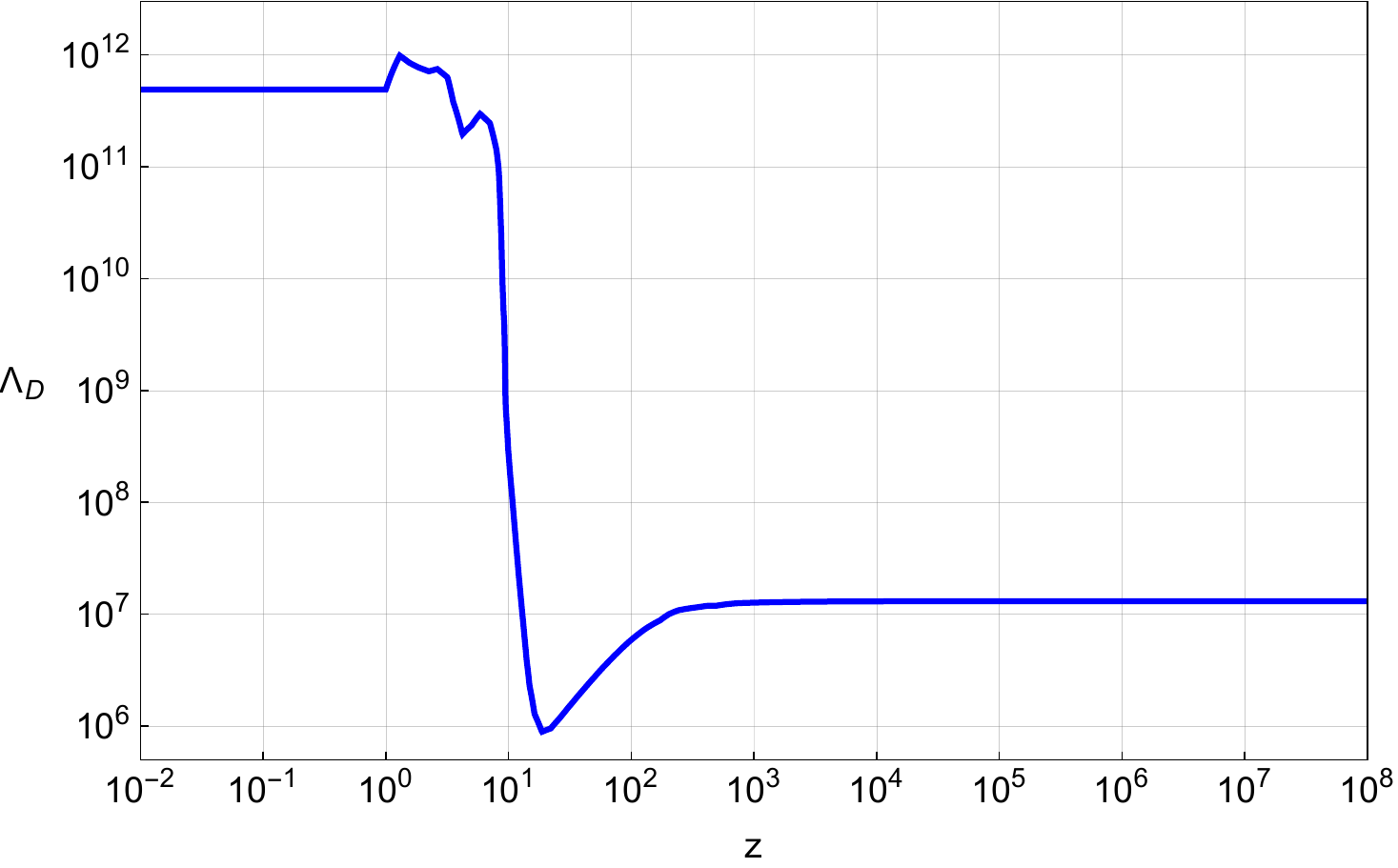}
\includegraphics[width=\columnwidth]{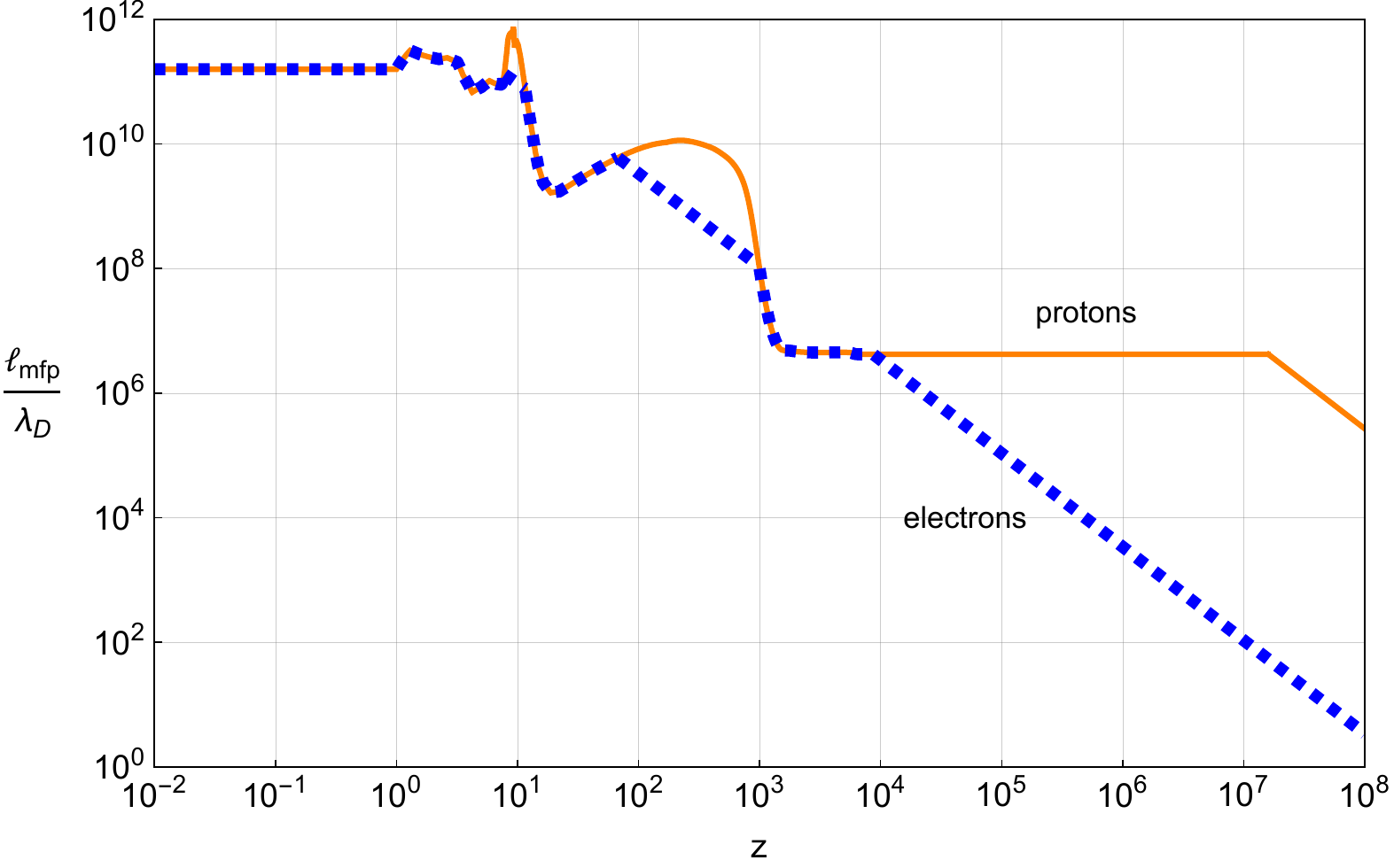}
\includegraphics[width=\columnwidth]{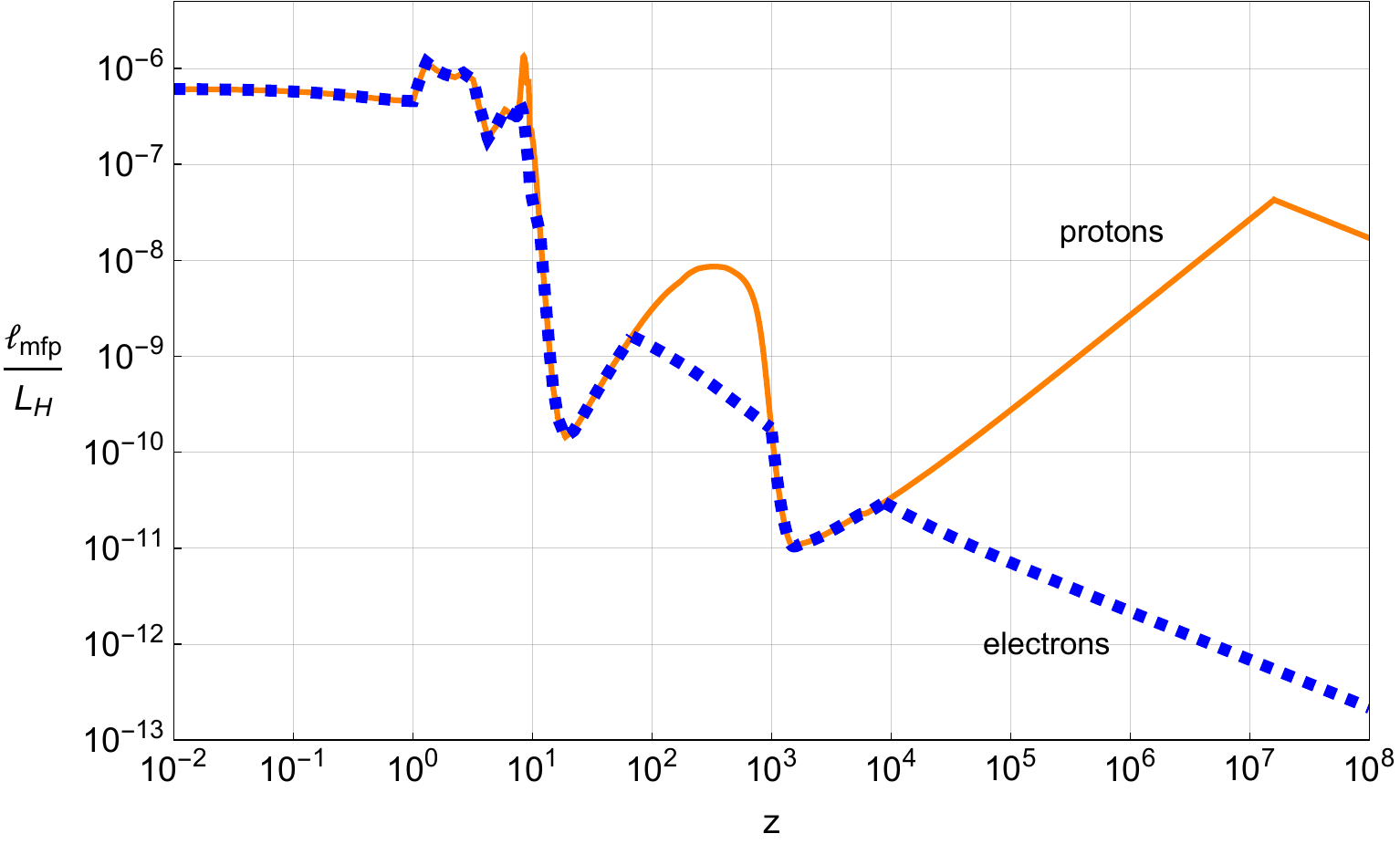}
    \caption{The universe is a collisional plasma at cosmological scales throughout its thermal history. 
    \emph{Upper panel:} plasma parameter as a function of redshift.
    \emph{Middle panel:} ratio of the mean free path of protons (orange, solid curve) and electrons (blue, dashed curve) to the Debye length as a function of redshift.
    \emph{Bottom panel:} ratio of the mean free path of protons (orange, solid curve) and electrons (blue, dashed curve) to the Hubble length as a function of redshift.} 
    \label{fig:plasma}
\end{figure}

As shown in Fig.~\ref{fig:plasma}, 
the universe is a collisional plasma at cosmological scales throughout its thermal history.
In the upper panel of Fig.~\ref{fig:plasma}, we show that
the plasma parameter 
\begin{eqnarray}
    \Lambda_D=\frac{4}{3}\pi \,n_b \,\lambda_D^3\,, ~~~~{\rm with}~\lambda_D=\sqrt{\frac{T_b}{4\pi e^2 n_b}}\,,
\end{eqnarray}
expressing the number of particles in a Debye sphere, is always much larger than one. 
In the middle panel,  we show the ratio of the electrons and protons mean free paths $\ell_{\rm mfp}=\langle v_{e,p}\rangle \tau_{e,p}$ (accounting for both Coulomb and Thomson interactions, see Sec.~\ref{subsec:sigmaev}) divided by the Debye length $\lambda_D$, and show that this is also always much larger than one. 
A collisional plasma is one for which the relevant scales satisfy $L\gg \ell_{\rm mfp}$.
In the bottom panel of Fig.~\ref{fig:plasma} we show the ratio of the electrons and protons mean free paths $\ell_{\rm mfp}$
and the Hubble length 
$L_H$, as a function of redshift. 
This quantity is always much smaller than one. 
In particular, focusing on the case of the voids today, we can appreciate that they can be treated as filled with a collisional plasma on scales $L\gtrsim 10^{-6} L_H\simeq 3\times 10^{-3} h^{-1}$ Mpc. 

\begin{figure}[b]\begin{center}
\includegraphics[width=\columnwidth]{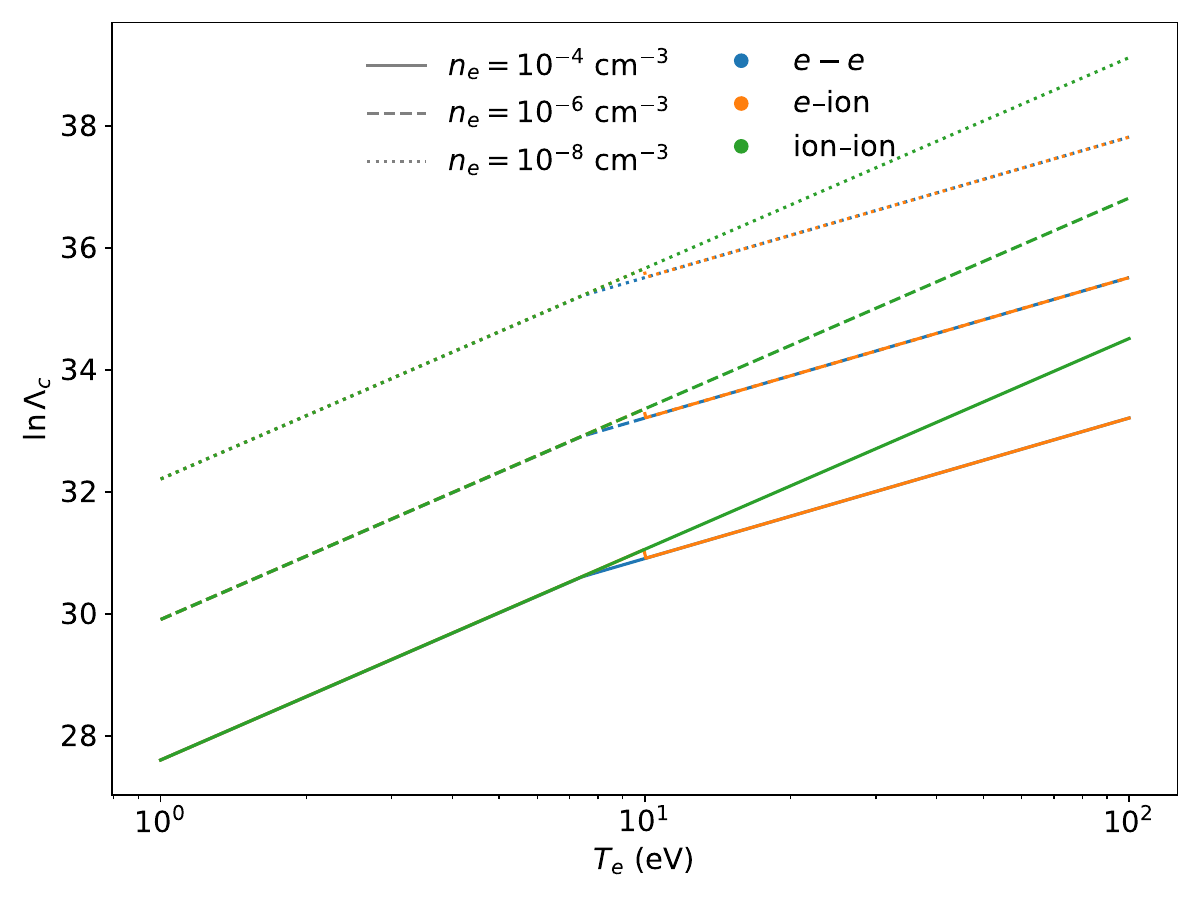}
\end{center}\caption[]{
Dependence of Coulomb logarithm with electron density, for various collisional processes shown in different colors. The number density along the x-axis goes from void to clusters to galactic environments
}\label{fig:coulomblog}\end{figure}

\section{Dependence of Coulomb logarithm on density and temperature}
\label{app:coulomblog}

There are three regimes depending on the types of collision taking place in the plasma
\begin{enumerate}
    \item For thermal electron-electron collisions \begin{equation}
\ln \Lambda_\mathrm{c}=23.5-\ln \left(n_e^{1 / 2} T_e^{-5 / 4}\right)-\left[10^{-5}+\left(\ln T_e-2\right)^2 / 16\right]^{1 / 2}.
\end{equation}

    \item For thermal electron-ion collisions \begin{equation}
\begin{array}{ll}
\ln \Lambda_\mathrm{c}=24-\ln \left(n_e^{1 / 2} T_e^{-1}\right) & 10 \mathrm{eV}<T_e \\
\ln \Lambda_\mathrm{c}=23-\ln \left(n_e^{1 / 2} T_e^{-3 / 2}\right) & T_e<10 \mathrm{eV}.
\end{array}
\end{equation}

    \item For thermal ion-ion collisions \begin{equation}
\ln \Lambda_\mathrm{c}=23-\ln \left(n_e^{1 / 2} T_i^{-3 / 2}\right).
\end{equation}

\end{enumerate}

From Fig.~\ref{fig:coulomblog}, we see that at $T=10~\text{eV}$, the behavior of Coulomb logarithm with respect to temperature for $e-$ion collision switches from that for ion-ion to that of $e-e$ at all densities of interest spanning clusters, filaments, and voids. Thus, at present day, for all practical purposes, we can refer to the part of the plot where all curves overlap at $\leq 10\eV$, without taking into account any additional source of heating in the IGM.

\end{document}